\documentclass[reprint,amsmath,amssymb,aps,twocolumn,floatfix]{revtex4-1}
\usepackage{graphicx} 
\usepackage{array}
\usepackage{amsmath}
\usepackage{dcolumn}
\usepackage{bm}
\usepackage[english]{babel}
\usepackage{epstopdf}

\begin{document}
\title{Frequency-control of protein translocation across an oscillating nanopore.}

\author{Fabio Cecconi}
\affiliation{Istituto dei Sistemi Complessi, CNR, 
                Rome, Italy} 
\author{Muhammad Adnan Shahzad} 
\author{Umberto Marini Bettolo Marconi} 
\affiliation{Dipartimento di Fisica, Universit\`a di Camerino, 
                Camerino, Italy.}
\author{Angelo Vulpiani}
\altaffiliation{Centro Linceo Interdisciplinare ``B. Segre'', 
Accademia dei Lincei, Rome, Italy.}
\affiliation{Dipartimento di Fisica, Universit\`a "Sapienza" di Roma, Italy}

\abstract{
The translocation of a Lipid Binding Protein (LBP) is studied using a phenomenological 
coarse-grained computational model that simplifies both chain and pore geometry.
We investigated via molecular dynamics the interplay between transport and unfolding 
in the presence of a nanopore whose section oscillates periodically in time with a
frequency $\omega$, a motion often referred to as radial breathing mode (RBM).
We found that the LPB when mechanically pulled into the vibrating nanopore exhibits a 
translocation dynamics that in some frequency range is accelerated and shows a frequency locking to the pore dynamics.
The main effect of pore vibrations is the suppression of stalling events of the translocation dynamics, 
hence, a proper frequency tuning allows both regularization and control of the overall transport process.
Finally, the interpretation of the simulation results is easily achieved by resorting to a first passage theory of elementary driven-diffusion processes.} \\
}
\maketitle
\section{Introduction}
Various biological and technological reasons require
to study the translocation of macromolecules across nanopores 
under conditions that vary cyclically in time.
Among them, we can mention, the recurrences imposed by metabolic cycles 
on the processes governing the transport of 
biopolymers across cellular compartments \cite{Principles}. 

A typical example can be found in the action of 
certain proteases (ClpXP) that, upon transforming the energy of ATP hydrolysis 
into mechanical force, unfolds and translocates polypeptides into the 
associated nanopores where they are eventually degraded.
The translocation occurs in cycles composed of a dwell phase, 
during which the polypeptide is at rest, and a burst phase, 
in which the polypeptide is pulled \cite{Maillard2011459}.  

Nanopores in thin and flexible membranes, like graphene layers, 
are not rigid and their thermal fluctuations may have a 
non-negligible impact on the translocation dynamics of long 
molecules \cite{grapheneNP_NModes}. 
According to normal mode (NMA) and to 
principal component (PCA) analysis, it is customary to 
decompose fluctuations of the membrane either in normal or principal 
modes to reveal the most important movements. 
Then, in a mechanical view of the system, the role of each mode 
can be studied separately and the analysis is restricted to those 
modes with largest contributions to the atomic mean square 
displacement (MSD).
In a nutshell, the procedure amounts to applying a periodic 
deformation (a mode) to the pore contour. 


On the experimental and technological side, time modulation of translocation 
processes can arise from spontaneous or induced variations in laboratory conditions.
For instance, the employ of alternating electrical sources finds applications 
in pulsed voltage driven experiments \cite{SlowDNA,harrell_Res,PolyMoby} or 
pulsed-field gel electrophoresis \cite{Pulsed_EPh}. 

Other laboratory experiments \cite{fanzio2012modulating,tuneableNP} proved that 
deformation of nanopores by an applied stress allows the control of DNA 
translocation speed. 
Upon this basis, a sequence of compressions and releases of nanopores is expected
to yield a cyclical behaviour on macromolecule transport.
This technique is a promising method for controlling translocation process by means
of a periodic modulation of mechanical stress and constitutes a viable 
alternative to the methods based on tuning: electrolyte salt concentration, viscosity 
or electrical voltages.

Apart from the obvious biological and technological interest, the theoretical 
interpretation of  translocation  experiments in time-modulated environments 
is particularly challenging as it involves different approaches of Statistical 
Physics, ranging from, biopolymer modelling, to transport theory, 
to methods of stochastic processes.  

Several computational and theoretical studies that addressed the effects of a modulated 
driving on translocation have mainly focused on simple unstructured polymers.
In this context, some authors considered periodic pulling fields 
\cite{anissilaJCP2012,Spagnolo_JCP2013,Fiasco2015}, others,   
instead, constant field and placed the modulation on 
the environment: nanochannels 
\cite{Golest2011,SRes1,SRes2} or solvent \cite{Golest2012}. 
Recently, the effects of concomitant time variations of field and channel have been 
theoretically analyzed in Ref.\cite{Ala-Nissila2015}.

The most relevant contribution emerging from these studies is the discovery of 
a noise induced phenomenology in polymer translocation akin to stochastic 
resonance (SR) \cite{StocRes} and {\em resonant activation} (RA) 
\cite{Doering,Astumian}, according to which the average 
translocation time as a function of the frequency of the external 
forcing presents a non-monotonic behaviour characterised by a 
sequence of minima and peaks 
\cite{Spagnolo_PhysBio010,anissilaJCP2012,Golest2011}. 
At it remarked by several authors, the RA can be observed in 
environments that undergo either oscillatory or random fluctuations.  

Inspired by these works, we set out to study the generic effect of 
a pure radial vibration of a cylindrical nanopore on the 
translocation properties of protein-like structures, by implementing a 
simple coarse-grained model that correctly describes secondary motives 
and compactness of the protein to be imported. 
In the following, we borrow the acronym RBM (radial breathing mode) from the 
carbon-nanotube literature \cite{RBM} for indicating the radial periodic 
expansion-contraction of the pore. 

We focus on a molecule belonging to Lipid Binding Protein (LBP) family
that share a simple barrel-like fold. Such proteins can reversibly and 
non-covalently associate with lipids, favouring the solubility of lipids in water and 
facilitating their transport between tissues. Regardless of its function, 
the LBP has been selected for its barrel topology that results in a clear 
sequential breaking of secondary motives under mechanical pulling by the C-terminus. 
Moreover, the barrel constitutes a symmetric and compact core which can easily give 
rise to stalled translocation dynamics when imported in a narrow pore.    

In this paper only steric-like interactions between the pore and 
the protein are taken into account; a RBM determines a modulation of 
the steric hindrance to protein passage that virtually resemble a cycle of 
"open-closed" pore states. 
Our primary purpose is understanding how a RBM modifies the RA mechanism 
when simple polymers are replaced by polypeptide chains with a well-defined compact 
geometrical structure.
Indeed, the natural tendency of proteins to fold into globular compact states is expected
to interfere with both entrance and translocation in nano-confined geometries leading
to an irregular transport behaviour.
The greater complexity with respect to linear polymers is ascribable to the 
following main reasons:
a) transport of proteins in narrow pores requires partial or full 
chain denaturation, as a consequence, unfolding and transport are often coupled. 
In the literature, this coupling is generally referred to as 
{\em co-translocational unfolding} \cite{Andricioaei,Bayley,grapheneNP,Chinappi,
cressiot2014focus};
b) the geometrical properties of protein chains is known to influence the 
translocation kinetics. Indeed, some structural elements or blocks, either for 
robustness or compactness, contribute to stall the process in dynamical intermediates,  
one is thus allowed to coin the term {\em structure-dependent translocation}; 
c) {\em multiple-strand translocation} occurs when 
a passing protein allocates simultaneously multiple strands inside the channel, 
in contrast to the single-file mode where the passage occurs strand by strand. 
The multiple-strand passage represents one of the main factors slowing down 
the translocation.

In this respect, it is natural to wonder how the scenario described in a), b) and c) 
modifies under pore RBM oscillations. 
In particular, three issues can be specifically addressed by our simplified mechanical 
model.  

The first concerns the general response of the LPB translocation dynamics to the pore 
mechanical action, to what extent the translocation and pore dynamics are resonant. 
Another issue refers to how certain pore vibrations might affect the presence and the 
impact of possible translocation intermediate 
states on the dynamics \cite{Andricioaei,bacci2012role,Bacci2013,grapheneNP}.  

Finally, we wonder if the RBM of the channel is able to trigger or 
accelerate the translocation dynamics in analogy 
with the results of Ref.~\cite{knots} on knotted proteins. 


We will start by analysing the LBP translocation across a static pore which has 
to be considered as the reference case. 
Simulations show that the translocation dynamics is characterised by a major stall 
event occurring when a last residue of secondary structure involving, strands 
S1 (segment 46-52), S2 (segment 56-63) and S3 (segment 67-72) reaches the pore entrance.  
A stall is the trapping of molecule conformations into on-pathway intermediate 
states that are considered long-lived when compared to the whole translocation
duration. Furthermore, 
the persistence of this block inside the pore leads also to translocation events 
that are not single-file. 

Then, we study how the LBP translocation gets modified when the pore undergoes 
RBM with frequency $\omega$. Three regimes are observed.
At low frequency, long stalls are not suppressed but their duration is reduced, 
overall, the translocation process remains slower than that of the static case.\\
At intermediate frequency, stalling events are significantly suppressed 
with a consequent speeding up of translocation with respect to the static pore.\\ 
Finally, in the high frequency regime, we find a further improvement in translocation 
efficiency accompanied by a modulation of the LPB dynamics with 
the pore vibration; a clear indication that pore oscillations couple to LPB transport 
dynamics. Beyond this range, the protein dynamics is no longer able to lock to the 
forcing applied by the pore. 
Even in the absence of an obvious locking between pore and protein dynamics, 
we observe, above a certain frequency threshold, a translocation speeding up 
with respect to the static pore, basically due to the mechanism of 
{\em stalling suppression}. 

The paper is organised as follows: sec.II we briefly summarise the computational 
model used to simulate LBP translocations in a narrow pore. 
In sec.III we illustrate and discuss the results in the case of a 
static pore and we extend the analysis to the fluctuating pore.

\section{Computational model
\label{sec:compumodel}}
We implemented a coarse-grained representation of both protein  
and pore, where the pore is simplified to a confining
channel with soft walls and the LBP chain, whose atomic coordinates 
are downloaded from the Protein Data Bank (pdbid: 2MM3 \cite{2MM3}),
is reduced to a sequence of point-like beads which spatially coincide 
with the C$_\alpha$ carbons of the protein backbone.  
Despite this great simplification, the protein-like nature of the LBP structure is 
preserved when modelling the intrachain interactions via a G\={o}-type force field 
proposed by Clementi et al. \cite{Clementi} that takes into account in a realistic 
manner the secondary-structure content (helices and beta-sheets) of a protein chain.   
This characteristic is crucial in the present study model, as we 
are interested in quantifying how the tendency of the macromolecule to maintain 
its globular native conformation reflects on the translocation dynamics.
  

In their approach, the force field acting on the $N$ beads is defined by four 
potential-energy terms: 
\begin{eqnarray*}
\Phi_{G\bar{o}} & = & \sum_{i=1}^{N-1}V_{p}(r_{i,i+1}) + 
\sum_{i=1}^{N-2}V_{\theta}(\theta_{i}) + \sum_{i=1}^{N-3}V_{\varphi}(\varphi_{i}) \\
       & + &  \sum_{i,j \geq i+3}V_{nb}(r_{ij}) 
\end{eqnarray*}
The peptide term, $V_{p}$, that enforces chain connectivity, 
is a stiff harmonic potential allowing only small oscillations of the 
bond lengths around their equilibrium values
$$
V_{p}(r_{i,i+1}) = \frac{k_{p}}{2}(r_{i,i+1}-R_{i,i+1})^{2}
$$
where $r_{i,i+1}$ and $R_{i,i+1}$ are the distances between beads 
$i$ and $j$ in the current and native conformation, respectively. 
The spring constant is $k_{p}=10^3 \,\epsilon/d_{p}^{2}$, with $\epsilon$ setting the
energy scale and $d_{p}=3.8\AA$ is the average distance between two 
consecutive residues.
 
Likewise the bending potential $V_{\theta}$
allows only small fluctuations of the bending angles $\theta_{i}$ around 
their native values $\Theta_{i}$
$$
V_{\theta}(\theta_{i}) = \frac{k_{\theta}}{2}(\theta_{i}-\Theta_{i})^{2}  
$$
where $k_{\theta}= 20 \,\epsilon \,$rad$^{-2}$. The native secondary structure
is primarily enforced by the dihedral potential $V_{\phi}$. Each dihedral 
angle, identified by four consecutive beads, contributes to the potential 
with the terms
$$
V_{\phi}(\phi_i) = k_{\varphi}^{(1)}[1 - \cos(\varphi_{i} - \Phi_{i})] +
k_{\varphi}^{(3)}[1 - \cos3(\varphi_{i} - \Phi_{i})]
$$
where $\Phi_{i}$ denotes the value of the $i$-th angle in the native 
structure,
$k_{\varphi}^{(1)} = \epsilon$ and  $k_{\varphi}^{(3)} =0.5\epsilon$.

Finally, the long-range potential $V_{nb}$ which favors the formation of the 
correct native tertiary structure by promoting attractive interactions
is the two-body function,   
\begin{eqnarray*}
V_{nb}(r_{ij}) = \epsilon 
\begin{cases} 
5\left(\dfrac{R_{ij}}{r_{ij}}\right)^{12} - 
6\left(\dfrac{R_{ij}}{r_{ij}}\right)^{10} & R_{ij} \leq R_c\\[10pt]
      \dfrac{10}{3}\;\left(\dfrac{\sigma}{r_{ij}}\right)^{12} & R_{ij} > R_c\;.
\end{cases}
\end{eqnarray*} 
Therefore, the interaction between aminoacids $i-j$ is attractive 
when their distance in the native structure, $R_{ij}$, is below a certain cutoff, 
$R_{c}=7.0\AA$ in this work, otherwise the aminoacids repel each other via a soft-core 
interaction with $\sigma=5\AA$. 
It means that the system gains energy as much as a pair of beads involved
in a native contact is close to its native configuration.

A unique parameter $\epsilon$ sets the energy scale of the force field
the other parameters introduced above are the typical ones used
in similar G\={o}-type approaches, see e.g. 
\cite{Clementi,hoang2000,cecconi2006testing}

A Langevin thermostated dynamics evolves the position ${\mathbf r}_{i}$ of 
the $i=1,\ldots,N$ aminoacids
\begin{equation}
M_{aa} \ddot{\mathbf r}_{i} = -\gamma \dot{\mathbf r}_{i} - 
\nabla_{r_i}\,(\Phi_{G\bar{o}} + V_{pore}) + {\mathbf F_N} + 
\mathbf Z_i\,.  
\label{eq:Langev}
\end{equation}
Where $M_{aa}$ denotes the average aminoacid mass of the protein chain,    
$\mathbf Z_i$ is a random force with zero average and correlation
$\langle Z_{i,\mu}(0) Z_{i,\nu}(t)\rangle = 2\gamma k_B T
\delta_{\mu,\nu} \delta(t)$, with $\mu,\nu = x,y,z$ and $k_B$ being the 
Boltzmann's constant.
$V_{pore}$ indicates the channel potential defined below and ${\mathbf F_N}$
is the constant pulling force, acting only on C-terminus (last bead),
that drives the chain into the nanopore.      

The simulation implements dimensionless quantities, such that 
energy is expressed in units $\epsilon$, masses in units $M_{aa}$,     
and length in units $\sigma$.  
Consequently, temperature, time and force are measured in units: 
$\epsilon/k_B$ and $t_u =\sigma\sqrt{M_{aa}/\epsilon}$, 
$f_u = \epsilon/\sigma$ respectively. 
The dynamics~\eqref{eq:Langev} is integrated 
via a stochastic leap-frog algorithm (\cite{leapfrog-SDE} p.251), 
with a time step $\Delta t = 0.0025\,t_u$ and $\gamma = 0.25/t_u$, at a temperature
$T = 0.3\,\epsilon/k_B$.
To convert the code units into physical ones, we simulated the thermal denaturation 
of the LBP structure with a set of equilibrium MD runs at increasing temperature. 
The data, combined and analysed via the multiple histogram method 
\cite{WHAM}, yielded a folding temperature $T_f = 0.72$ which 
corresponds to an experimental denaturation temperature of $348^o$K \cite{IFABP}. 
The matching between simulated and experimental temperature sets the energy scale 
to the value $\epsilon \simeq 1.0\,$Kcal/mol. 
In addition, since the total molecular mass of the $127$ aminoacids of LPB is 
$14.26\,$KDa, 
we obtain the average mass $M_{aa} \simeq 1.86\times 10^{-25}\,$Kg, hence, 
the unit time scale turns to be $t_u \simeq 2.6\,$ps and force $f_u = 13.73\,$pN.  

\begin{figure}[h]
\centering
  \includegraphics[height=5cm]{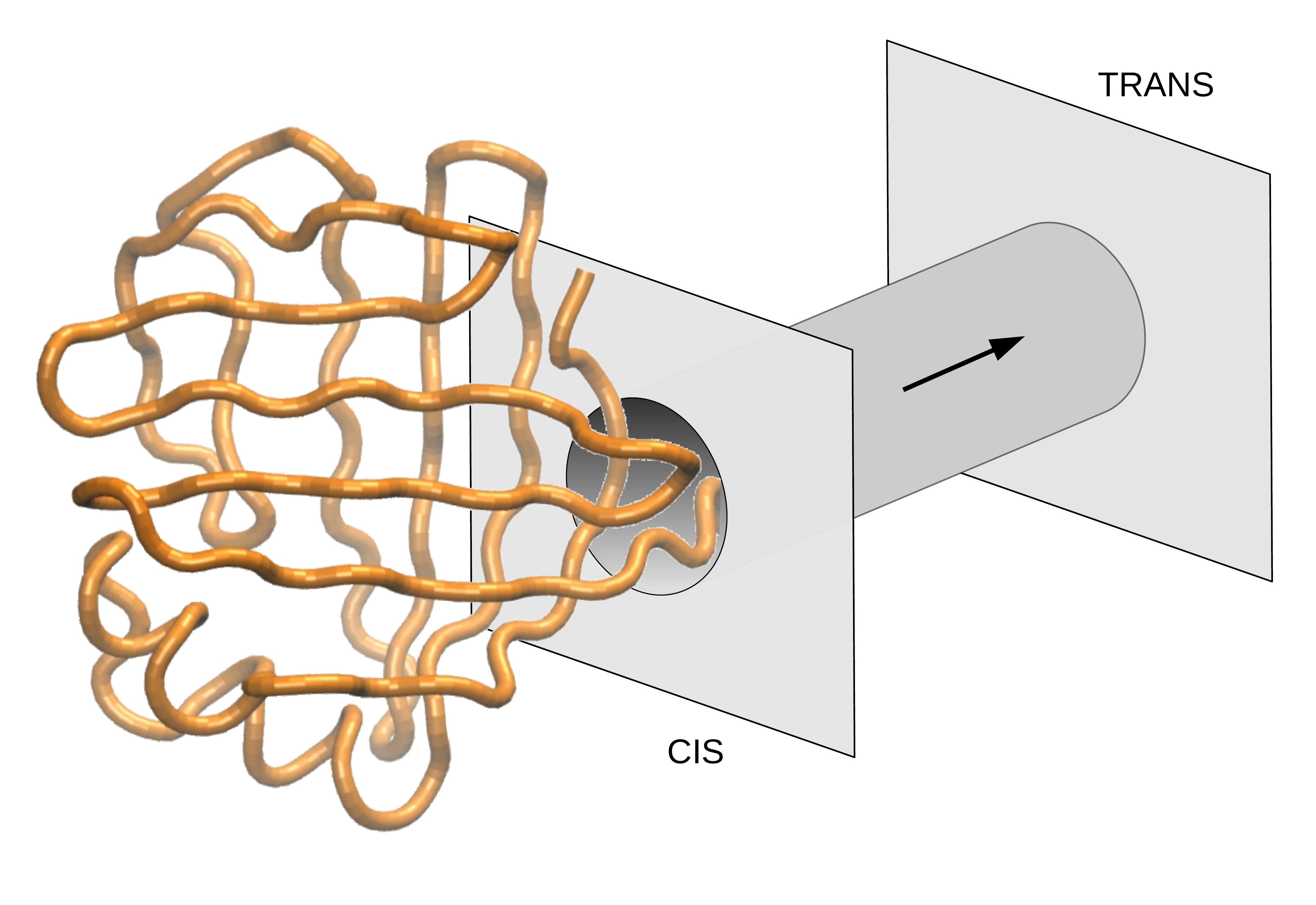}
\caption{\label{fig:pwalls} Cartoon of the setup employed in our molecular 
dynamic simulations. The LPB backbone chain extracted from the file 2MM3.pdb 
\cite{2MM3} is prepared at the channel entrance,  
vertical walls represent the membrane 
whose impenetrability is guaranteed by the repulsive force
$\{F_w(x),0,0\}$ normal to the walls. 
The arrow indicates the pulling direction from CIS to TRANS side.} 
\end{figure}

We model the nanopore through which the protein is transported into as a confining 
cylindrical region centered along the $x$-axis (translocation direction) 
with length $L$ and time dependent radius $R_p(t)$, Fig.\ref{fig:pwalls}.  
The confinement is obtained via a potential of cylindrical symmetry simulating a 
hole in a soft wall 
\begin{equation}
V_p(\mathbf r,t) = \epsilon
\begin{cases} 
(\rho^2 - 1)^{\alpha} & \mbox{if~} \rho > 1 \\
0                     & \mbox{if~} \rho \le 1
\end{cases} 
\end{equation}
where $\rho = (y^2 + z^2)/R^2_p(t)$. 
The parameter $\alpha$ controls the stiffness of the confinement. 
We are interested in the case where  
$R_p(t)=R_0[1+\delta\sin(\omega t)]$ 
oscillates around the static value $R_p$, with a sinusoidal 
law of frequency $\omega$, amplitude $\delta$.  
A repulsive force, $F_w(x)$, orthogonal 
to planes $x=0$, $x=L$ and vanishing for $y^2 + z^2 < R_p^2$,    
mimics the presence of the impenetrable 
membrane where the pore is inserted in 
$$
F_w(x) = 
\begin{cases} 
-\dfrac{e^{\lambda x}}{x + c}             & x \le 0 \\
                        0                 & 0 < x < L \\ 
  \dfrac{e^{-\lambda (x-L)}}{x - L + c}   & x \ge L   
\end{cases} 
$$ 
with $c=10^{-4}\AA$ being a regularisation cutoff to avoid overflow 
near the walls. 
In this work, we choose $\alpha=3$ and amplitude $\delta=0.3$. 
The pore length $L=100\AA$ and radius $R_0=10\AA$ are taken from 
$\alpha$HL structural data \cite{aHL}. Since $R_0$ is smaller than the gyration 
radius of the folded LBP structure, full translocations 
imply partial or complete unfolding. 

For facilitating the entrance of the chain into the nanopore,  
an inert linker of five extra beads was added to the N-terminus of 
the LPB, this linker extends the free tail protruding from the 
globule that has to be pulled.

The importing mechanism that drives the protein into the pore 
is simplified to a constant pulling force $(F,0,0)$ acting only 
on the N-terminus bead ($\mathbf r_N$)  
in such a way that the pulled terminus is constrained to slide 
along the pore axis for all time, i.e.,  $y_N(t) = z_N(t) = 0$.  

\section{Results}
We import the LBP from left to right inside the pore, and simulations 
are run until the whole chain lies outside the channel, on the cis-side. 

The initial conformation of each translocation run is 
obtained by equilibrating the chain at code temperature 
$T = 0.3$ ($T \simeq 150^o$K) and $F=0$, while  
the pulling terminus is kept at the position ${x_N = -1, y_N = 0, z_N=0}$, 
near the entrance of the static pore ($\omega=\delta=0$).

Once the translocation run is completed, a new run is restarted from a different 
thermalised initial condition, the procedure is repeated until a robust statistics of 
translocation events is collected. 

Even for a coarse-grained description of the protein dynamics, the conformation space 
is still very high-dimensional to allow a concise representation of the 
translocation. 
It is thus convenient to ``project'' the system trajectories
onto an effective (or collective) coordinate that is a function 
of the aminoacid positions.
 
A suitable choice suggested by Polson et al.~\cite{Polson}
is the collective variable 
\begin{equation}
Q = \frac{1}{N} \sum_{i=1}^N g(x_i)\,
\label{eq:Q}
\end{equation}
defined by the piecewise function 
$$
g(x) = 
\begin{cases}
0     &  \mbox{if~}  x < 0 \\
x/L   &  \mbox{if~}  0 \leq x < L \\
1     &  \mbox{if~}  x \geq 1    
\end{cases}
$$
The value $Q = 0$ corresponds to the 
 whole protein on the cis-side, while $Q=1$
to  a successful translocation. 

Along with $Q$, we also monitor the number of LBP residues that 
during the translocation lie on the cis-side of the pore: 
\begin{equation}
N_{cis}(t) = N - \sum_{i=1}^{N} \Theta(x_i)\,,
\label{eq:Ncis}
\end{equation} 
$\Theta(s)$ being the unitary step function. 
This quantity, during a translocation event,  
starts from the maximal value $N_{cis} = N$ and decreases to zero. 
Even if $N_{cis}$ is not a good progress coordinate, as the state 
$N_{cis} = 0$ does not entail yet completed translocation events,  
it allows locating the position of the stalling points along the chain 
because stalls manifest as plateaus in the $N_{cis}$ time course.  

An important physical quantity of translocation processes is the 
passage time, i.e. the time the molecule takes to cross the pore.
If we assume that the LBP is prepared at $t=0$ on the pore entrance (CIS-side),
the first-passage time is the first time at which the molecule lies outside 
the pore exit (TRANS-side), and it can be easily defined in terms of $Q$:
$$ 
t_{\mathrm{FP}} = \min_{0<t \le T_w} \{t\;|\; Q(t)=1\}
$$ 
where $T_w$ is the observation time window. As the statistics of $t_{\mathrm{FP}}$ 
can be easily measured in experiments, it is important to predict the dependence of 
$t_{\mathrm{FP}}$ on the system parameters such as the: chain length, type of driving 
force, pore fluctuations etc.

A reference theoretical framework for the statistical analysis of $t_{\mathrm{FP}}$ 
assumes $Q$ to be a random process governed by a 
driven Brownian motion \cite{Lubensky,Berezhkovskii2003,MuthuBook,Ammenti2009}
\begin{equation}
\dot{Q} = \mu_0 F + \sqrt{2 D_0}\;\xi(t)
\label{eq:DriDiff}
\end{equation}
where $\mu_0 F$ accounts for average drift determined by the pulling mechanism 
and $\sqrt{2 D_0}\;\xi(t)$ embodies both thermal and environment fluctuations. 
A successful translocation event requires that a trajectory of $Q(t)$ 
is released at $Q(0) = 0$ (cis) and terminated when $Q(t)=1$, (trans).

The statistics of $t_{\mathrm{FP}}$ associated with Eq.\eqref{eq:DriDiff} is well 
known since the works by Schr\"{o}dinger and Smoluchovski \cite{Schr1915,Smoluch,
RednerGuide} and it is characterized by the Inverse Gaussian distribution 
\cite{InvGauss}.
However for practical purposes explained below, it is more convenient to employ the 
cumulative distribution function (CDF) of the Inverse Gaussian, that reads
\begin{equation*}
F_{IG}(t) = \dfrac{1}{2} 
\bigg\{
    \phi\left(\dfrac{L - \mu_0 F t}{2\sqrt{D_0 t}}\right) + 
    \mbox{e}^{\mu_0 F L/D_0} \phi\left(\dfrac{L + \mu_0 F t}{2\sqrt{D_0 t}}\right)
\bigg\}
\label{eq:IG-CDF}
\end{equation*}
where  
$\phi(u) = \mbox{Erfc}(u)$ is the complementary error function \cite{Arfken}.

\subsection{Static pore}
We begin our analysis from the case of a static pore, $\delta = \omega=0$,
when the translocation dynamics of the LPB presents the interesting feature of 
a stalled event, for which the translocation progress is not uniform in time, 
but gets jammed when certain chain segments approach the pore entrance. 
These stalling points are associated with specific LBP conformations that 
are particularly difficult to be unravelled. To some extent, they behave 
as ``temporary knots'' of the 
chain \cite{knots,knot2,Makarov} contributing to a remarkable transport slowdown.

Stalling events that are particularly persistent are to be considered 
intermediates of translocation, as they are statistically robust to imprint an 
unmistakable multistep signature on observables in experiments 
\cite{Talaga,Bayley,clpx2011,clpx2013,nivala} and simulations 
\cite{Andricioaei,bacci2012role,Bacci2013,grapheneNP}
\begin{figure}
\centering
  \includegraphics[height=5cm]{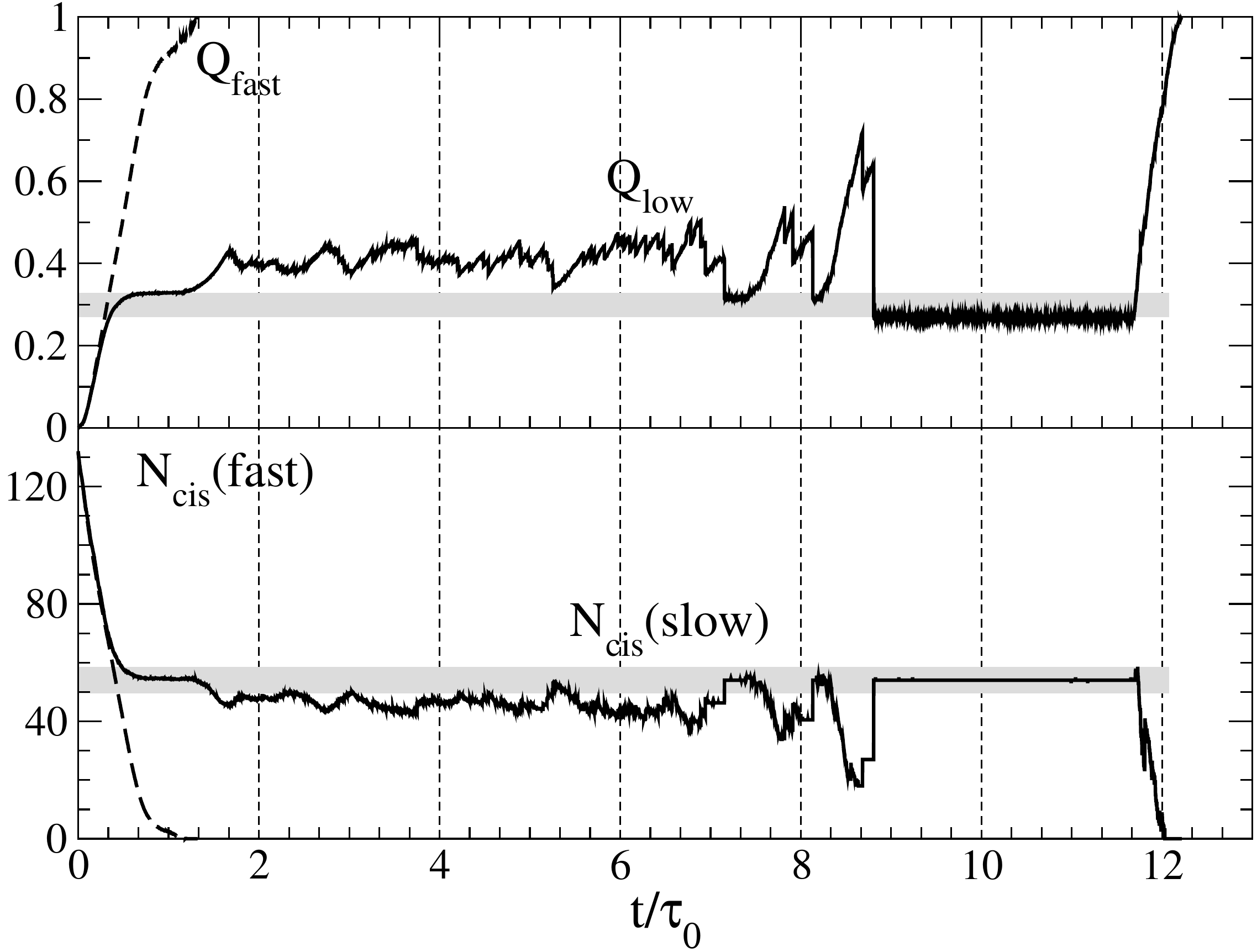}
\caption{\label{fig:stalls} Time course of 
$\langle N_{cis}(t)\rangle_{S,F}$ and $\langle Q(t)\rangle_{S,F}$ 
where subscripts denote that the averages are restricted to 
the ensemble of fast and slow translocations for a total of 
$M=3600$ independent runs.     
Slow translocations are characterised by intermediates (stalls) 
that contribute with a noisy plateau to the mean time behaviour (gray bands).
In the lower panel, the value $\langle N_{cis}(t) \rangle_S \simeq 54$ 
indicates a stalling point of the LPB due to a persistence of the 
intra-chain interactions that stabilise the block 
of strands S1 (4--12), S2 (46--53) and S3 (46--53), 
see Fig.~\ref{fig:topoLBP}.}
\end{figure}
We run $M=3600$ successful LPB translocations, each with a
duration $t_{m}$, leading to a mean translocation time 
$\tau_0 = \sum_{m=1}^{M} t_{m}/M = 8279.2\,t_u = 21525.9\,$ps.
In the following, the time will be rescaled with $\tau_{0}$.
In such simulations, the signature of a stalled dynamics turns to be evident 
by looking at the time behaviour of the two averages  
$\langle Q(t)\rangle $ and $\langle N_{cis}(t)\rangle$ taken over an ensemble 
 Fig.\ref{fig:stalls}.  
Both indicators show early variations which are then followed by  
long stationary phases before reaching their absorbing state ($N_{cis}=0, Q=1$).
To analyse separately the short and long time behaviours, it is convenient to split 
$\langle Q(t)\rangle $ and $\langle N_{cis}(t)\rangle$ in fast and slow components.
We define slow translocations those which are completed in a time 
$t > T_{cut} = 1.08\times 10^{4} t_u$.
While the fast components (dashed lines) saturate soon to the expected values in a 
monotonic way, the slow components follow the fast ones for a while, then  
deviate toward a flat stationary noisy behaviour corresponding
to the stalled state with $N_{cis} \simeq 54$. 

We compute the fraction of time the LPB chain spends in a state with a 
given $N_{cis}$, this quantity is defined by the histogram  
$$
H(N_{cis}) = \dfrac{1}{T_M} 
\sum_{m=1}^{M} \int_{0}^{t_m} \!dt\;\delta[N_{cis} - N_{cis}(t)]
$$
where the sum runs over the $M$ translocations 
and $T_M = \sum_{m=1}^{M} t_{m} = M\tau_0$.

\begin{figure}
  \centering
\includegraphics[height=5cm]{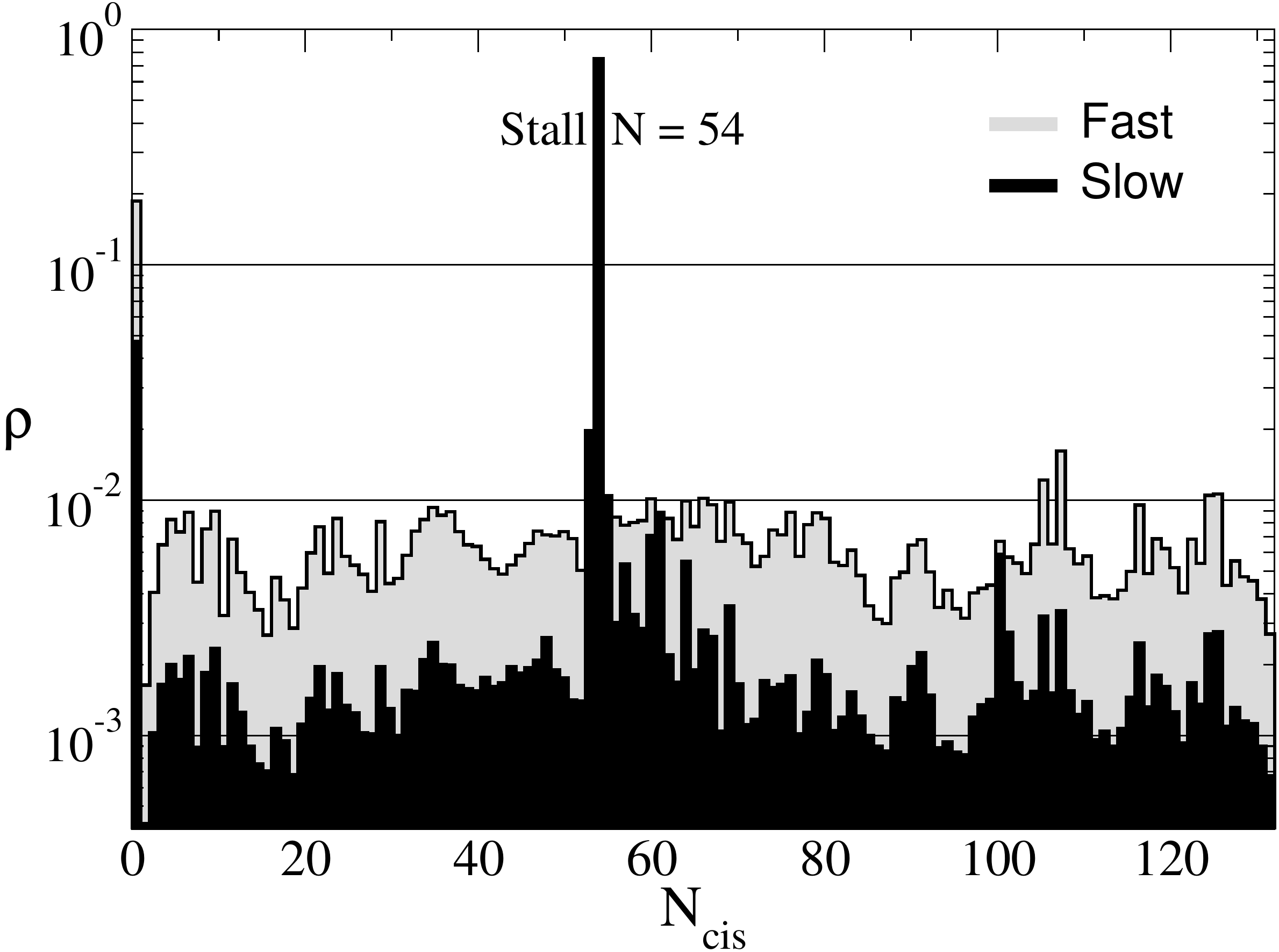}
\caption{\label{fig:freq}
Visiting frequency (histogram) of states $N_{cis}$ during 
each translocation run. The occurrence frequency is split into 
``fast'' and ``slow'' components as it is computed over fast and slow 
translocation sets.} 
\end{figure}

\begin{figure}
  \centering 
 \includegraphics[height=5cm]{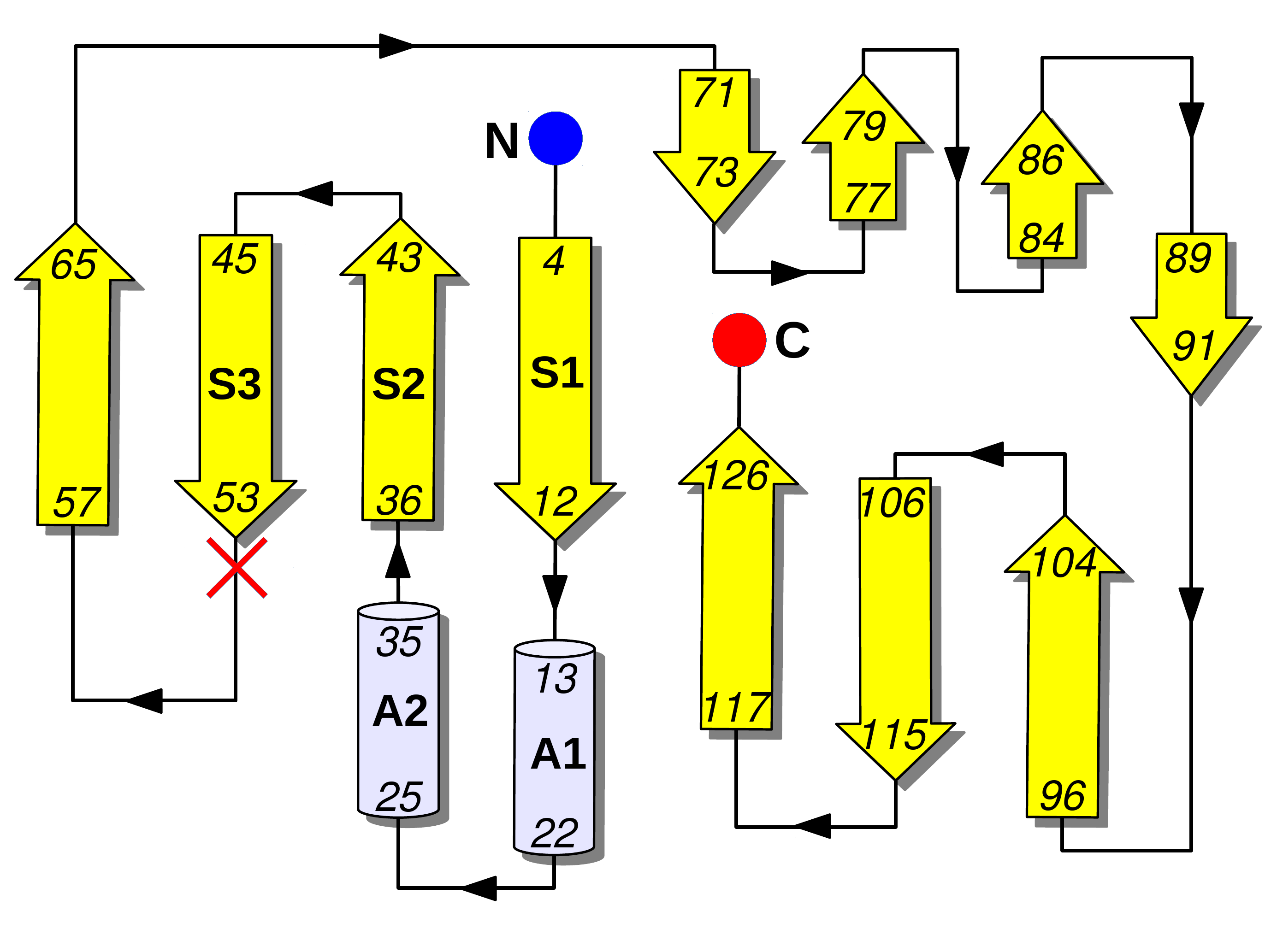}
\caption{\label{fig:topoLBP}
Topology of the LBP native structure, arrows and cylinders represent
strands and helices, respectively. 
The red cross marks the stalling site, 54, located
just after the end of strand S3 (46--53), that  
with strands S1 (4--12), S2 (36--43) forms the core  
of a stable block, weakly involving helices A2 (13--22), A2 (25--35).
The protein is pulled from the C-terminus (Red dot).}   
\end{figure}

\begin{figure}
 \centering
 \includegraphics[height=5cm]{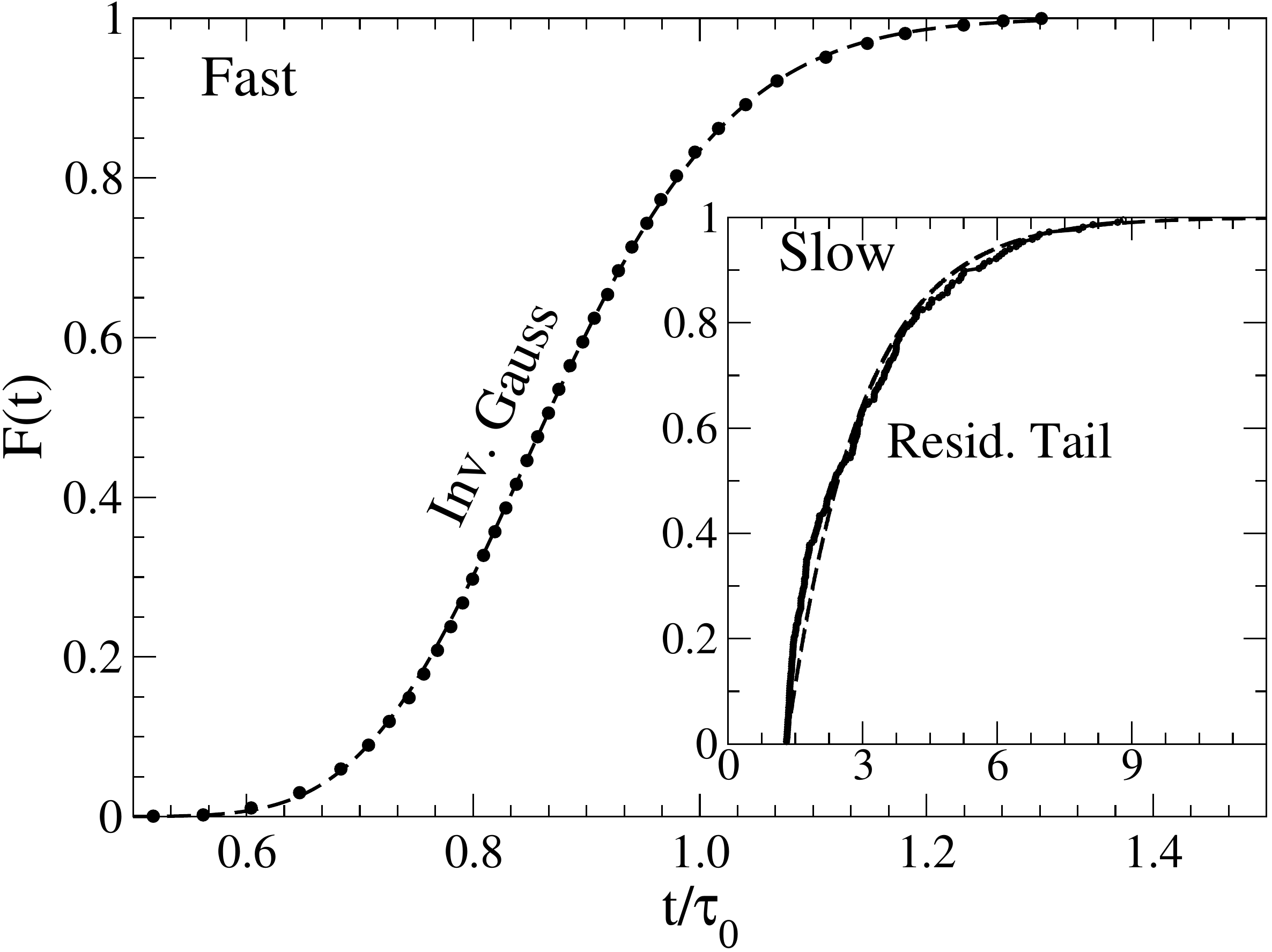}
\caption{\label{fig:invGauss}
Empirical CDF, $F_{M}(t)$, of translocation times associated to fast events
(dots) compared with $F_{IG}(t)$ (dashed line).
Parameters $D_0 = 1.56\times 10^{-6}$ and $\mu_0 = 6.9\times 10^{-5}$ 
are obtained by the Maximum 
Likelihood values of the Inverse Gaussian, Eqs.\eqref{eq:parma}. 
Inset: Empirical CDF of the translocation times of slow events 
characterized by stalls. Points are the simulation data and dashed line 
is the CDF of an exponential distribution with $\lambda = 7.3\times 10^{-5}$ 
and $T_{cut} \simeq 1.08\times 10^{4}$.}
\end{figure}

Likewise, we split $H(N_{cis})$ into fast and slow components,
Fig.\ref{fig:freq}. 
The histogram of fast events is practically flat 
indicating that each chain conformation is uniformly visited. 
On the contrary, the slow component presents a narrow and pronounced peak
emerging from the background in $N_{cis} = 54$, confirming that 
the LBP chain spends a relevant amount of time in the state 
with the bead $54$ at the pore entrance. 
With reference to the LBP native structure topology, Fig.~\ref{fig:topoLBP}, 
one can deduce that stalling at site $54$ lies just after
the end of S3 (segment 46--53), suggesting that S3 along with two 
other strands, S1 (segment 4--12) and S2 (segment 36--43) 
form a block that is structurally robust, likely involving also 
the participations of the two helices 
A2 (13--22), A2 (25--35).  
The persistence of such a block causes the jamming of the protein 
moreover, it often squeezes into the pore and translocates 
as a single unit. 
In other translocation runs, instead, the block breaks down 
allowing a true single-file passage of the molecule.   
To complete the characterization of the LPB translocation dynamics, 
we analyzed the statistics of translocation time by computing the 
empirical CDF over a sample of $M$ successful translocations 
occurred at times $\{t_1,\ldots,t_M\}$, 
$$
F_{M}(t) = \dfrac{1}{M} \sum_{k=1}^M \theta(t-t_k)\;, 
$$ 
where $\theta(s)$ denotes the unitary step function. 
The advantage of the CDF over the histogram lies in its 
independence of binning, so it is not affected by 
the chosen discretization.
Fig.\ref{fig:invGauss} displays the comparison between the expected 
and the empirical CDF for LPB translocations.
The values of parameters $D_0$ and $\mu_0$ have been obtained 
from the maximum likelihood estimation (MLE),   
\begin{equation}
\mu_0 = \dfrac{1}{F \langle t \rangle}\quad,\quad
D_0 = \frac{1}{2}\bigg(
\left\langle\dfrac{1}{t}\right\rangle - \dfrac{1}{\langle t \rangle} 
\bigg)     
\label{eq:parma}
\end{equation}
where angular brackets stand for the arithmetic average over 
$M$ independent realizations 
$\langle x \rangle = (x_1 + x_2\cdots + x_M)/M$.   
The inset of Fig.\ref{fig:invGauss} shows the empirical CDF of 
slow translocations that is consistent with the  
CDF = $1 - \exp[-\lambda(t - T_{cut})]$ of an exponential probabilistic law.

In conclusion, the comparison of the CDFs indicates 
fast translocations contribute to the Inverse Gaussian bulk of the time 
distribution, whereas, few slow translocations are responsible for the  
slow exponential decay of the long-time tail.
 
In the next section, we study how the above transport scenario 
characterised by stalling points is modified when the pore section 
undergoes periodic fluctuations.

\subsection{Oscillating pore \label{sec:flupore}}
We repeated the translocation runs at different frequencies $\omega$
of the radius modulation to assess how the pore dynamics affects both 
the LBP mechanical denaturation and the subsequent transport. 
In particular, it is interesting to analyse the robustness of 
RA scenario \cite{Spagnolo_PhysBio010,Park-Sung,
anissilaJCP2012,Golest2011} when translocation dynamics 
is affected by the presence of extreme events like stalls.
 
\begin{figure}
  \centering 
\includegraphics[height=6.2cm]{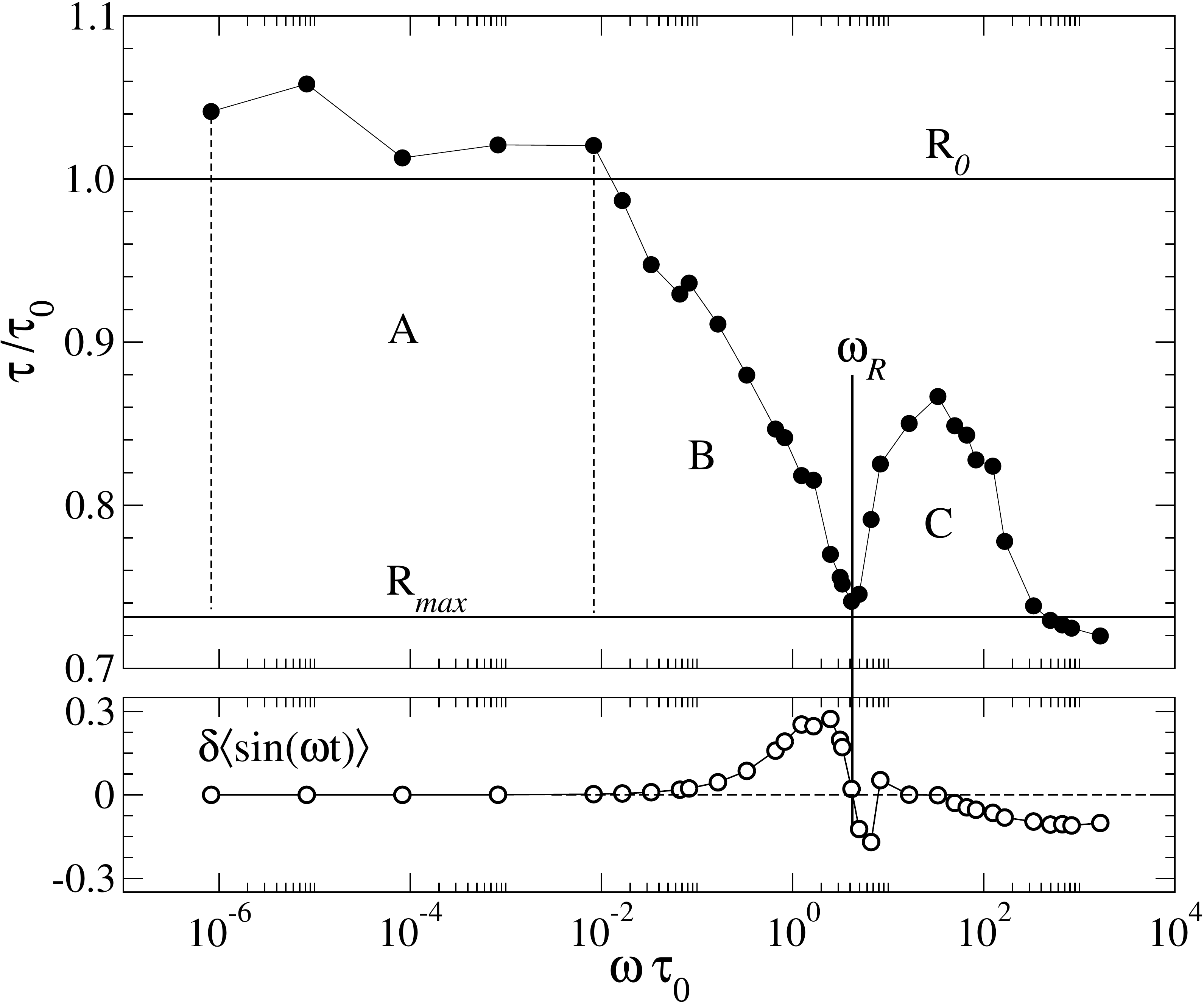} 
\caption{\label{fig:t_vs_omega} Upper panel:
Dependence of the mean translocation time $\tau$ on the
frequency at a force $F=2.0\,f_u = 27.46\,$pN. Rescaling with $\tau_0$ 
makes data dimensionless. Three translocations regimes can be 
identified corresponding to regions A, B, C in the $\omega$-range. 
A part from region A, the RBM of the pore generally 
reduces the translocation time and the transport across the nanopore 
is speeded up by oscillations. 
Lower panel shows the behaviour of Eq.\eqref{eq:osc} as a function of the rescaled 
$\omega$.  
The first value at which $\langle \sin(\omega\tau) \rangle = 0$, 
marked by the thick vertical line,
identifies the resonant frequency $\omega_R$ at which $\tau(\omega)$ 
attains its minimum.}
\end{figure}

We begin by plotting in Fig.\ref{fig:t_vs_omega} the dependence of the mean 
translocation time $\tau$ on the frequency of pore vibration. 
Data are rescaled with the static mean translocation time, 
$\tau_{0} = 21525.9\,$ps, {\em i.e.} $\tau/\tau_0$ vs. $\omega\tau_0$.
The horizontal lines mark the translocation time for 
the static pore with radius: $R_{max}=13\AA$, $R_0=10\AA$, for a comparison.
  
We observe different translocation regimes (labelled A,B,C) 
resulting in a non monotonic behaviour of translocation time versus 
the forcing frequency, in analogy with translocation of structureless polymers 
\cite{Spagnolo_JCP2013,anissilaJCP2012,Golest2011}. 

In region A ($\omega \tau_0 < 10^{-2}$), 
the average translocation time is close to, but larger than, 
the static value, $\tau(\omega) > \tau_0$, indicating a moderate slowing down 
of translocations with respect to the still pore. 
In the intermediate regime ($10^{-2} < \omega\tau < \omega_R \tau_0 $),   
region B, $\tau(\omega) < \tau_0$ and it decreases with $\omega$. 
In this range, vibrations speed up the transport dynamics with respect to 
the static case moreover, the acceleration improves by increasing $\omega$
till reaches an optimal value at $\omega_R \tau_0$.  

Finally, in regime C, $\tau(\omega)$ 
attains a maximum which yet lies below $\tau_0$, whereby 
translocations result to be still improved by the RBM dynamics. 

To verify that the plot of Fig.\ref{fig:t_vs_omega} is 
consistent with RA \cite{Doering}, we studied   
the two-state dynamics defined by the Langevin equation
\begin{equation}
\dot{Q} = \mu_0 F - \mu_0 \dfrac{\partial V}{\partial Q} + \sqrt{2D_0} \xi(t)\,,
\label{eq:two_state}
\end{equation}
obtained by adding to Eq.\eqref{eq:DriDiff} a force term derived from 
the time dependent potential
\begin{equation}
V(Q,t) = \dfrac{V_0\,f(\omega t)}{4}\;Q^2(Q - 1)^2
\label{eq:biwell}
\end{equation}
which represents a ``caricature'' of a translocation free-energy landscape
where, presumably, a barrier separates two minima: $Q = 0$ (cis) and $Q = 1$ (trans). 
The amplitude $V_0$ is multiplied by 
\begin{equation}
f(\omega t) = 1 + h\sin(\omega t + \phi)
\label{eq:modula}
\end{equation}
to account for barrier oscillations $V_b(t) = V_0\,f(\omega t)/64$.
The phase $\phi$ is an extra parameter necessary to fit the model to the 
pore RBM: opening translates into barrier lowering, while, pore shrinking  
corresponds to increasing the barrier. 
 
\begin{figure}
 \centering
  \includegraphics[height=5cm]{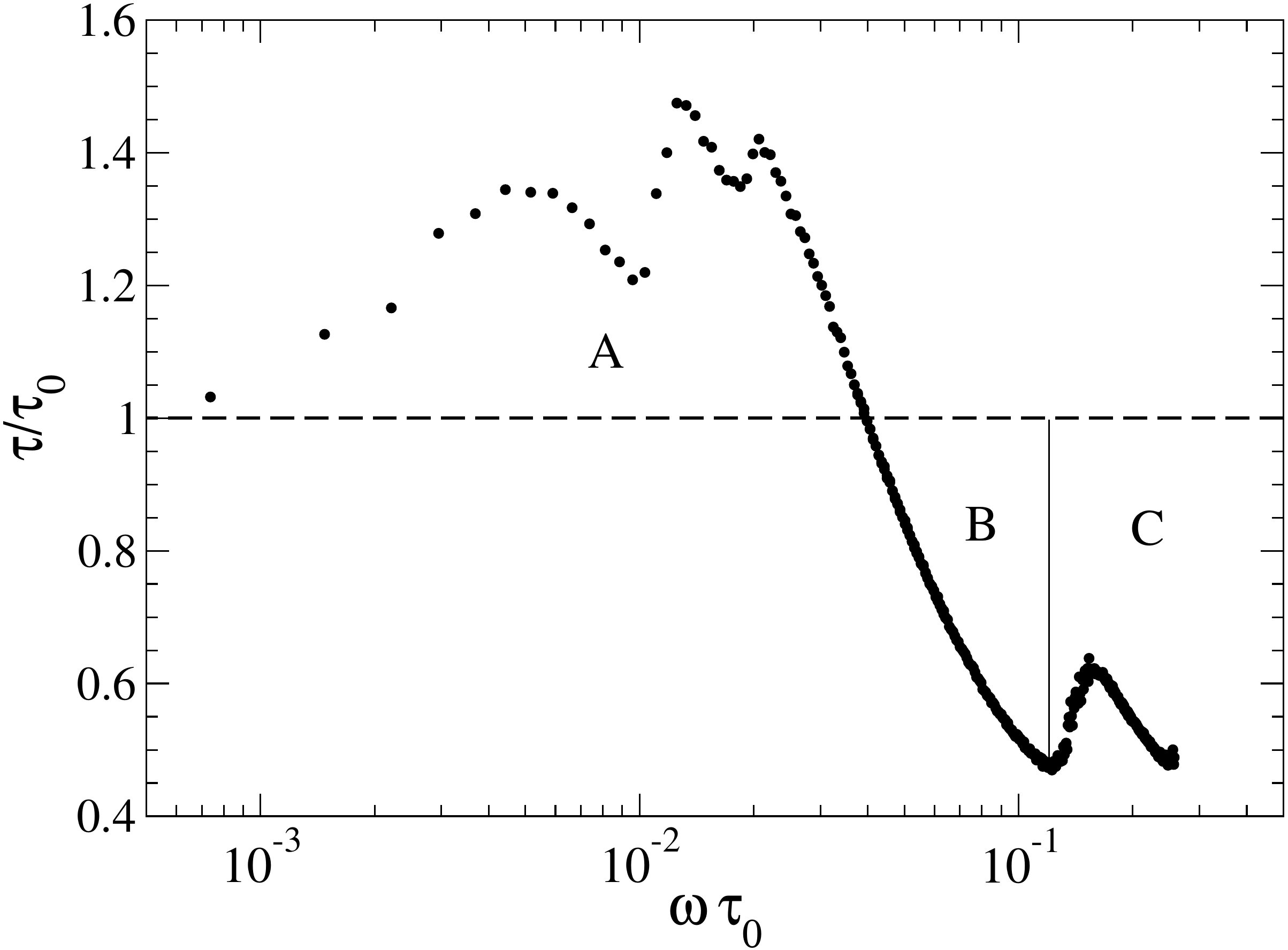} 
\caption{\label{fig:two_state} Two-state model with a fluctuating barrier.
First arrival time of trajectories started at $Q=0$ and absorbed at $Q=1$,
obtained by a numerical integration of Eq.\eqref{eq:two_state} 
with parameters $D_0=1$, $F=0.001$, $V_0=8\times 64$, $h=0.3$, 
and $\phi=0$.
Plot and frequency regions A,B,C are consistent with Fig.\ref{fig:t_vs_omega}.
}
\end{figure}

We integrated numerically Eq.\eqref{eq:two_state} via a second order stochastic 
Runge-Kutta algorithm \cite{StocRK} and computed the 
average first-arrival time to the state 
$Q=1$ from $Q(0)=0$, over a set of $10^4$ trajectories. 
Fig.\ref{fig:two_state} shows the mean first-arrival time 
as a function of $\omega$. 
The qualitative similarity between plots in Fig.\ref{fig:two_state} and 
Fig.\ref{fig:t_vs_omega} suggests that RA is verified and that 
translocation of the LPB across a vibrating channel can be idealised as a 
transition to an absorbing state over an oscillating barrier. 
Both plots in agreement with RA exhibit a minimum of $\tau(\omega)$ at 
a certain 
``resonant'' or ``optimal'' frequency $\omega_{R}$, separating regions A and B, 
at which the fastest translocations are expected to be observed.
 
A simple physical argument suggests that 
LPB translocations are greatly favoured as long as they are completed in a time 
interval $\Delta t$ such that the pore stays ``open'': 
$R_0[1 + \delta\sin(\omega \Delta t)] > R_0$,
corresponding to the inequality, $0 <\omega \Delta t <\pi$. 
Therefore, the optimal frequency $\omega_{R}$ at which translocations are 
faster is bounded in the range,
$\omega_0/2 < \omega_{R} < \pi/\tau(R_{max})$,
where $\tau(R_{max})$ denotes the mean translocation for a static pore
at maximal radius $R_{max} = R_0(1+\delta)$, indicated by the 
lowest horizontal line in Fig.\ref{fig:t_vs_omega}.
To verify this conclusion, in the lower panel of the same figure, 
we report the average 
\begin{equation}
\delta \langle \sin(\omega t) \rangle = 
\frac{1}{M} \sum_{i=1}^M \frac{R(t_i) - R_0}{R_0}\;,  
\label{eq:osc}
\end{equation} 
over $M$ translocation times $t_i$,     
estimating the typical radius oscillation at the end of translocations 
at each  $\omega$. 
It is apparent that by following the bold vertical line, that the first 
zero of $\langle\sin(\omega t)\rangle$ coincides with the resonant 
frequency $\omega_R$ at which $\tau(\omega)$ attains its minimum.
This confirms that at the resonant condition, $\omega_R \tau \simeq \pi$,
luckiest translocations occurs in the half-cycle of the RBM, in which the pore 
offers {\em minimal hindrance} to the transport.   
  
It is instructive to gain further insight into the physics of RA 
by adapting to our case the phenomenological approach to 
stochastic resonance 
by McNamara and Wiesenfeld \cite{theory_of_SR}. 
It amounts to writing a rate equation    
for the activated kinetics of model \eqref{eq:two_state}  
with the help of the well-known Kramers formula \cite{kramers1940}
$$
k = \dfrac{\Omega_0 \Omega_b}{2\pi \gamma} \exp\{-\beta V_b\}\;.
$$ 
Where $\gamma$ is the solvent viscosity, $\Omega_0, \Omega_b$, the angular 
frequencies (curvatures) at the bottom and at the top of the barrier
of the tilted potential, $W(Q,t) = V(Q,t) - F Q$.    
As it is shown in Appendix \ref{app:rate}, Kramers theory   
gives, for a weak external field $F$, the rate expression 
\begin{equation}
k(\omega t) \simeq a e^{-\epsilon \sin(\omega t + \phi)}[1+c\sin(\omega t + \phi)];
\qquad (\epsilon = \beta V_0 h/64) 
\label{eq:kramers}
\end{equation}
to the first order in $F/V_0$.  
The constant $c$ and the factor $a$ are defined in the appendix.

Eq.\eqref{eq:kramers} represents the lowest term of a Kramers escape-rate modulated 
by a unimodal potential vibration, high order terms in the expansion contribute 
with higher harmonics.    
This analytical approach remains physically meaningful 
as long as barrier oscillations are not 
too fast (adiabatic regime) with respect to the relaxation dynamics in the well:
the adiabatic regime requires $\omega \ll \Omega_0$, a condition certainly 
verified by the case in Fig.\ref{fig:two_state}.  

The rate-theory in appendix \ref{app:rate} shows that Eq.\eqref{eq:kramers} 
leads to the analytical expression 
\begin{equation}
\tau(\omega) = \dfrac{{\displaystyle
\int_{0}^{2\pi}}\!\!dx\;\exp\{-Y(x)/\omega\}} 
{\omega\big[1-\exp(\,-2\pi\;k_{\mathrm{eff}}/\omega\,)\big]} 
\label{eq:tau_res}
\end{equation}
for the average translocation time, with 
$$
Y(x) = \int_{0}^x\!\!du\;k(u)\;.
$$
being the integral of the rate \eqref{eq:kramers}
and 
\begin{equation}
k_{\mathrm{eff}} =  \dfrac{1}{T} \int_{0}^{T}\!\!dt\; k(\omega t) = 
\dfrac{Y(2\pi)}{2\pi}
\end{equation}
being the average rate over a period of vibration. 
The resonant frequency $\omega_R$ is the minimum of $\tau(\omega)$. 
As discussed in Appendix \ref{app:rate}, this minimum is close to the value 
$\omega_*  \pi\;k_{\mathrm{eff}}$ at which the argument of the exponential 
at denominator in Eq.\eqref{eq:tau_res} equals $1/2$. 

At low frequency, the escape from the barrier is basically determined by 
the frozen value of $k(0)$ (barrier height) that is selected by the initial condition of the dynamics. 
Whereas at high frequency, the escape is determined by the average rate (barrier)
$k_{\mathrm{eff}}$. The resonant minimum basically separates these two regimes. 

In summary, the basic condition for emergence of RA is an escape 
process modulated by a time periodic rate. It is reasonable to assume 
that a similar situation occurs in the LBP translocation. 
    
\begin{figure}
\centering
  \includegraphics[height=5cm]{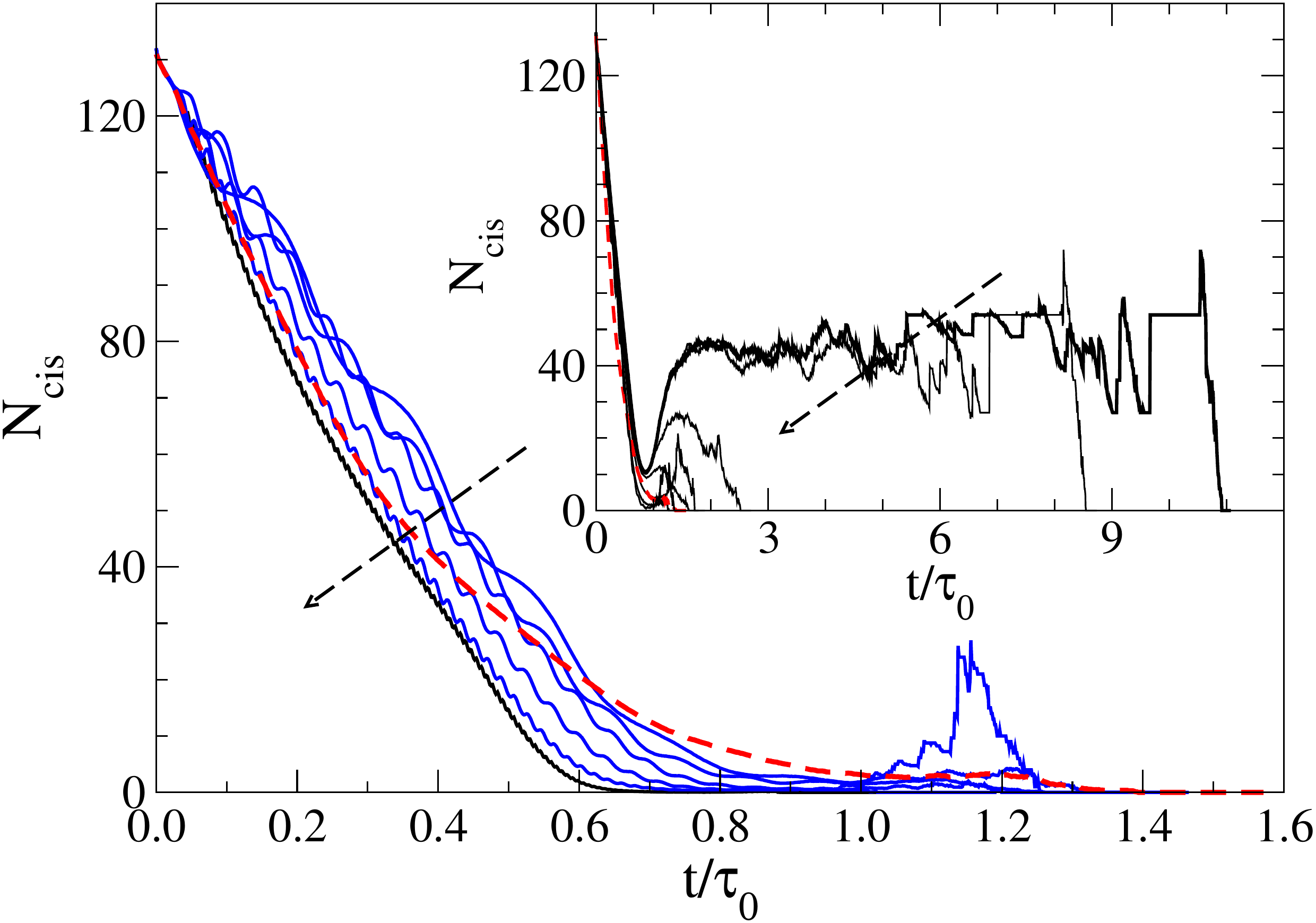}
\caption{\label{fig:kill_stall}
Main panel: modulated evolutions of $\langle N_{cis}(t) \rangle$ 
with the pore frequency as a consequence of the transmission of pore oscillations to 
translocation dynamics. This oscillating decay of $\langle N_{cis}(t) \rangle$ 
is well distinguishable only in the frequency range coincident with region C of 
Fig.\ref{fig:t_vs_omega}. 
The modulation is soon lost for frequencies just below region C, see 
the steady decay of the red-dashed curve, and it becomes imperceptible
at frequencies just above region C, see black-thick curve. 
Inset shows the stall suppression in translocations at pulling 
$F = 27.46\,$pN ($F=2.0$ code units) when increasing the oscillation frequency 
of the pore section, dashed arrows indicate the increase of $\omega$. 
It is apparent how the plateau length of $\langle N_{cis} \rangle$ reduces when $\omega$ 
increases, till vanishing when $\omega \to \omega_0$.  
} 
\end{figure}

Now it interesting to investigate the effect of the channel 
fluctuations on the persistence of the stalling events.
This can be achieved by measuring how much the trajectories of $N_{cis}$  
experiences the influence of the pore frequency. 
Fig.\ref{fig:kill_stall} reports the time course of the average $N_{cis}$ over 
$3600$ runs, for different values of $\omega$. 
We recall that $N_{cis}$ is particularly useful for identifying 
the stalled dynamics. 

From main panel of Fig.\ref{fig:kill_stall}, we observe that in 
the frequency range C (as defined in Fig.\ref{fig:t_vs_omega}), 
pore vibrations transfer to the translocation dynamics, 
indeed $\langle N_{cis}(t)\rangle$ 
develops an oscillating decay to zero with the pore frequency. 
The locking between pore and protein pulled-dynamics is expected,  
because if the pore is maximally closed, the protein dynamics is hindered 
and temporary stalled. 
In that condition, $N_{cis}$ statistically assumes  
the same value, leading to equally spaced peaks in the plots.  
However, the decay without oscillations shown by the red-dashed curve, 
obtained at a frequency just below the region C, proves    
that such a frequency locking is restricted to the frequency 
region C. 
The oscillation of the thick-black curve, corresponding to a 
frequency just above region C, is almost imperceptible because 
the protein dynamics is becoming not responding to such 
fast pore oscillations.

However, it should be remarked that the absence of an evident 
frequency locking does not imply a 
translocation dynamics which is not sensitive to the forcing 
applied by the pore. 
Indeed, the plots of $\langle N_{cis}(t)\rangle$ in the inset of 
Fig.\ref{fig:kill_stall} show 
that the duration of the stalls soon reduces as the pore vibrates even at low frequency,
and by increasing $\omega$ the duration is further decreased till vanishing above 
a certain frequency threshold.
We stress that time shortening of stalls does not necessarily imply an average 
speeding up of translocations, in particular, region A of Fig.\ref{fig:t_vs_omega} is 
just characterised by translocations with $\tau(\omega) > \tau_0$ notwithstanding the 
stall depression. 
If stalls are regarded as extreme events, 
their reduction is crucial in order to regularise the translocation dynamics 
according to the principle that suppression of extreme events generally  
makes a process more predictable and controllable \cite{Sornette}.    

The data suggest that a specific cycle of expansions and compressions of the channel 
may either control or even facilitate the translocation of proteins across it. 
The result can be summarised by the statement: 
``tuned RBM of a nanopore catalyses pulled translocation of globular 
proteins''.  
As shown in Refs.\cite{knots,Andricioaei}, an analogous catalytic effect can be 
achieved by setting the modulation on the pulling force that unfolds and 
translocates polypeptide chains and proteins. 

\begin{figure}
\centering
  \includegraphics[height=5cm]{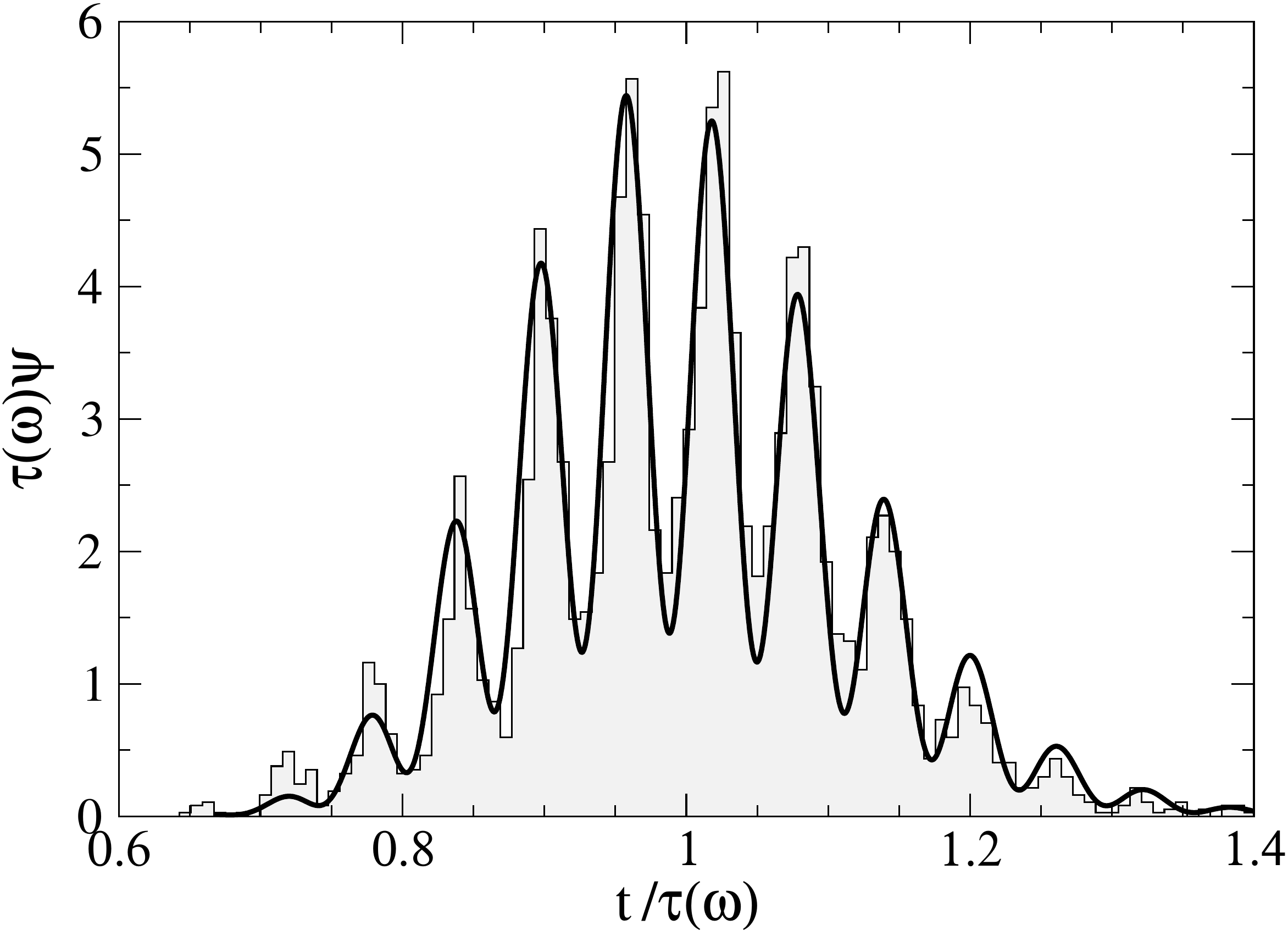}
\caption{\label{fig:fpt_peak} Comparison of the PdF of the translocation time for 
$\omega \sim 20 \omega_0$ to the corresponding analytical PdF \eqref{eq:pdf_time}
with parameters $\mu_0=7.33\times 10^{-5}$, $D_0 \sim 1.012\times 10^{-6}$, 
$h = 0.60$ and $\phi = -10$.} 
\end{figure}

In the region C (Fig.\ref{fig:t_vs_omega}), we observe that  
PdF of translocation time develops a multi-peaked structure 
reflecting the pore cycles, see Fig.\ref{fig:fpt_peak}.
The minima and maxima of the PdF correspond to a maximally open and closed state 
of the pore respectively. Outside the region C this PdF modulation either vanishes 
or becomes undetectable.  

Again a simple approach that can explain the multi-peaked structure of the 
translocation time PdF is bases on a First Passage Theory (FPTh) for 
a biased diffusion of $Q$ described by the Smoluchowski equation
\begin{equation}
\frac{\partial P}{\partial t} = 
-\mu_0 F \frac{\partial P}{\partial Q} + 
D_0 f(\omega t)\frac{\partial^2 P}{\partial Q^2}  
\label{eq:FPQ}
\end{equation}
where $f(\omega t)$ is defined in Eq.\eqref{eq:modula}, 
$\mu_0$ and $D_0$ denote static mobility and diffusivity 
respectively. 
In Eq.\eqref{eq:FPQ} instead of taking a periodic pulling force,
as done in other contexts \cite{anissilaJCP2012,Fiasco2015}, 
we preferred to consider a systematic drift $\mu_0 F$, 
while shifting the modulation to the diffusion coefficient 
$D_0 f(\omega t)$.   
This approach is more consistent with the 
coarse-grained molecular model implemented in our simulations and 
described in sect.\ref{sec:compumodel}, 
where the sinusoidal oscillation of the pore applies cyclic 
transversal compressions on the passing chain leading to a 
kind of ``freezing'' on the transversal degrees of 
freedom.  This has been roughly taken into account by a noise with an 
oscillating variance. 
The solution to Eq.\eqref{eq:FPQ} is specified by initial  
$P(Q,0) = \delta(Q)$ and boundary conditions, 
\begin{eqnarray*}
J(0,t) & = & \mu_0 F P(0,t) - 
D_0 f(\omega t)\dfrac{\partial P}{\partial_Q}\bigg|_0 = 0 \\ 
P(1,t) & = & 0 \,; 
\label{eq:boundary}
\end{eqnarray*}
the first equation is a no-flux condition which guarantees that $Q(t)$ 
cannot be less than zero, by definition. The second prescribes that trajectories 
are absorbed as soon as $Q(t)=1$. 
 
The fundamental quantity in the FPTh is the survival probability of $Q(t)$ 
$$
S(t) = \int_{0}^{1}\!\! dQ P(Q,t) 
$$ 
where $P(Q,t)$ is the solution of Eq.\eqref{eq:FPQ} satisfying both boundary and 
initial conditions.
$S(t)$ is the probability that at time $t$ the process $Q(t)$ is not yet absorbed  
by the boundary $Q=1$, accordingly, $1 - S(t)$ is the probability that 
$Q(t)$ exits $[0,1]$. Hence, the exit time distribution is $\psi(t) = d(1-S(t))/dt$, 
that is  
$$
\psi(t) = -\dfrac{dS}{dt} = -\int_{0}^{1}\!\!dQ \dfrac{\partial P(Q,t)}{\partial t}
$$
Using Eq.\eqref{eq:FPQ}, we obtain that $\psi(t)$ is related to  
probability flux evaluated at the boundary, $Q=1$,
\begin{equation}
\psi(t) = J(1,t) = D_0 f(\omega t) \dfrac{\partial P}{\partial Q}\bigg|_{Q=1} 
\label{eq:psi}
\end{equation}
therefore, the final result reads
\begin{equation}
\psi(t) = \frac{1 + h\cos(\omega t+\phi)}{\sqrt{4 D_0 \Delta^3(t)}} 
\exp\bigg\{-\frac{(1 - \mu_0 F t)^2}{4 D_0 \Delta(t)}\bigg\}\;.
\label{eq:pdf_time}
\end{equation}
where $\Delta(t) = t - (h/\omega) [\cos(\omega t+\phi)-\cos(\phi)]$ is 
obtained by integrating $f(\omega t)$ over time, (see Eq.\eqref{eq:Delta}).
The derivation of this theoretical distribution is outlined in Appendix \ref{app:Toy}
by using the {\em method of images} to fulfil the boundary conditions. However, it 
is important to warn that formula \eqref{eq:pdf_time} constitutes 
only a reasonable approximation of the true solution, indeed as it discussed by 
Molini et al. \cite{molini2011} and in Appendix \ref{app:Toy}, 
the image method to work, when applied to Smoluchowski equations with time 
dependent coefficients, requires a rigorous proportionality between drift and diffusion; 
a condition which is not verified in Eq.\eqref{eq:FPQ}. 
In addition, we assumed the further simplification of strong enough drift that 
soon pushes the trajectories away from the $Q=0$-boundary, so that the no-flux 
condition is automatically implemented. 

Despite the approximation, formula \eqref{eq:pdf_time} can be considered 
a good fitting model, that, upon tuning the parameters $\mu_0,D_0,h,\phi$, 
is able to reproduce and explain quite naturally the essential features of the 
simulated PdF, including the peculiar peaked structure as shown in 
Fig.\ref{fig:fpt_peak}, where the function \eqref{eq:pdf_time} fits 
well the histogram of LBP translocation times.

\section{Conclusions}
We investigated the translocation process of a protein in the family of 
Lipid Binding Proteins across a nanopore via a coarse-grained molecular
dynamic simulations that simplify both pore and chain.
In our phenomenological model, the protein is described as a chain of beads 
interacting via a G\={o}-like force field which is known to guarantee the correct 
formation of the secondary structure by rewarding those interactions that stabilise the 
geometry of the native state.     
The presence of a constant driving force mimics the average effect of the 
biological importing mechanism into a nanopore (cylinder) whose cross section varies 
periodically in time simulating the effect of a radial breathing mode (RBM) 
induced by a cyclically varying environment.

Our study differs from previous ones \cite{Spagnolo_JCP2013,anissilaJCP2012,Fiasco2015,
Golest2011,SRes1,SRes2,Golest2012} that focused on bead-spring polymers, 
for it investigates the translocation of a protein-like chain. The translocation of 
proteins is known to strongly deviates from that of polymers as their 
compactness presents much more resistance to the passage through narrow paths.  
This important feature, generally known as ``structure-dependent translocation'', 
makes the transport of proteins in nanopores a complex phenomenon still difficult to 
be both modelled and predicted. 

The Lipid Binding Protein (LBP) does not make an exception. Indeed, 
our MD simulations of its pulled translocation into a static pore, 
performed by the coarse-grained model, exhibit the typical intermittency 
of a process that is dominated by few extreme stalled events. More specifically, 
the chain gets temporarily stuck in metastable conformations that are hardly 
unravelled and depend on the LBP's arrangement in its own native state.  

We repeated the same simulations with a pore undergoing a radial vibration
(radial breathing mode (RBM)) to study the dependence of the average 
translocation time $\tau$ on the RBM frequency $\omega$.  
The comparison with the static case proved that the RBM reduces the duration of 
stalling events until it makes them disappear above a certain frequency threshold.

It is interesting to note that there exists a low frequency range, where 
the translocation process is slowed down by the RBM of the pore, despite a 
reduction of the stalling periods. 
The suppression of stalling duration, even if does not always bring to an  
accelerated transport, is crucial to ``regularise'' the process
by suppressing extreme events. 

In other regimes, a frequency locking occurs between the RBM and translocation 
dynamics; the translocation observable develops oscillations with the pore frequency and
the distributions of translocation time show a succession of peaks strictly 
reflecting such a locking.

\section{Acknowledgements}
The authors warmly thank A. Cavagna, R. Larciprete, M. Chinappi 
and L. Pilozzi
for very useful discussions and their valuable suggestions.

\appendix{} 
\section{Rate equation \label{app:rate}}
In this appendix we derive the escape rate \eqref{eq:tau_res}
by applying the Kramers' theory to the tilted bistable potential \eqref{eq:biwell}   
\begin{equation}
W(Q,t) = \frac{A(t)}{4} Q^2(Q-1)^2 - F Q\;. 
\label{eq:wpot}
\end{equation}
According to Kramers \cite{kramers1940}, 
the rate at which a Brownian dynamics leaves the left well of $W(Q,t)$ 
upon crossing the barrier reads
$$
k = \dfrac{\sqrt{W''(Q_0)\;|W''(Q_b)|}}{2\pi \gamma} \exp\{-\beta V_b\}.
$$
where $\beta = (k_B T)^{-1}$, $\gamma$ is the solvent viscosity, 
$W''(Q_0)$ and $W''(Q_b)$ are the second derivatives of the 
potential \eqref{eq:wpot} evaluated at the bottom of the well $Q_0$ and at the top 
of the barrier $Q_b$, respectively.  
Even in this simple framework, the rate formula 
becomes quite involved for a full analytical approach. 
The expression simplifies considerably in the limit of a weak field by retaining terms  
to the first order in $F/A$. The weak field shifts the well and barrier from the unperturbed
positions $Q_0 = 0, Q_b = 1/2$  to  $Q_0 \simeq 2 (F/A)$, $Q_b \simeq 1/2 - 4(F/A)$, moreover 
it decreases the barrier height from $ V_b = A/64  \to  A/64 - 0.5 (F/A)$. 
A further expansion of $k$ in $F/A$ yields the result
\begin{equation}
k(\omega t) \sim a \dfrac{1 + c \sin(\omega t +\phi)}{\tau_0} 
e^{-\epsilon\sin(\omega t + \phi)} 
\label{eq:ktime}   
\end{equation}
with constants $c = h/(1 - 6 F/V_0)$, $\epsilon = \beta V_0 h/64$, and 
$$
\tau_0 = 
\dfrac{4 \pi\;\gamma\;\sqrt{2}}{V_0 - 6 F}\; e^{\beta(V_0/64 - F/2)}   
$$
being the static escape time in the Kramers approximation. 
As a  consequence of barrier oscillations, the rate $k(\omega t)$ turns to be a 
periodic functions with period $T=2\pi/\omega$.  
The factor $a = \tau_0^{-1} \exp[\epsilon\sin(\phi)]/[1+c\sin(\phi)]$ ensures that
the limit $\omega\to 0$ recovers the static value $k(0) = 1/\tau_0$.
   
The kinetics of barrier crossing in the oscillating potential 
\eqref{eq:biwell} can be described by the rate equation   
\begin{equation}
\dfrac{dS}{dt} = -k(\omega t) S\,,
\label{eq:rateP}
\end{equation}
for the probability $S(t)$ that at time $t$ the process still occupies $Q_0$ well, 
$S(t)$ is also called the survival probability of $Q_0$-state.  
In the formulation of our problem, Eq.\eqref{eq:rateP} contains only the loss 
contribution, as the molecule is removed after each successful translocation 
and re-injected from the CIS-side (impossibility of back-transitions). 

Integration of Eq.\eqref{eq:rateP}, with initial condition $S(0) = 1$, leads to 
the solution
$$
S(t) = \exp\bigg\{-\int_{0}^t\!du\;k(\omega u)\bigg\}\,, 
$$ 
which describes a decay to zero as translocation proceeds.
The quantity $1 - S(t)$ is the probability that a molecule has crossed the boundary 
at time $t$ and it is removed, so the distribution of exit times is given by 
$\psi(t) = -S'(t)$, accordingly 
the mean exit (translocation) time is soon obtained from $\psi(t)$ and reads, 
$$
\tau(\omega) = \int_{0}^{\infty}\!\!dt\;S(t) = 
\int_{0}^{\infty}\!\dfrac{dt}{\omega}\;
\exp\bigg\{-\int_{0}^t\!\dfrac{du}{\omega}\;k(u)\bigg\}\,, 
$$
after an integration by part and a change of variables.
 
To take advantage of the periodicity of $k(u)$, 
the integral can be split in a series of integrals over the 
periods, corresponding to intervals $[2\pi n,2\pi n+2\pi]$,   
$$
\tau(\omega) = \dfrac{1}{\omega}\sum_{n=0}^{\infty}\;\int_{0}^{2\pi}\!\!d\xi\;  
\exp\bigg\{-\dfrac{1}{\omega}\int_{0}^{2\pi n+\xi}\!\!du\;k(u)\bigg\}
$$
where we applied the change of variable $t = 2\pi n + \xi$. 
Thanks to periodicity, we can refold the integration onto the 
cell $[0,2\pi]$ and the argument of the exponential is recast as
$$
n \int_{0}^{2\pi}\!\!du\;k(u) + \int_{0}^{\xi}\!\!du\;k(u)    
$$
with a final result 
$$
\tau(\omega) = \dfrac{1}{\omega}\sum_{n=0}^{\infty} 
e^{-n Y(2\pi)/\omega} \int_{0}^{2\pi}\!\!dt\;
\exp\bigg\{-\dfrac{1}{\omega}\int_{0}^t\!\!du\;k(u)\bigg\}. 
$$
According to the formula of a geometric-series sum, 
it becomes 
\begin{equation}
\tau(\omega) = \dfrac{{\displaystyle
\int_{0}^{2\pi}}\!\!dx\;\exp\{-Y(x)/\omega\}} 
{\omega\big(1-\exp[-Y(2\pi)/\omega\,]\big)} 
\label{eq:tauRA}
\end{equation}
where the function $Y(x)$ is the integral of the rate \eqref{eq:ktime}   
$$
Y(x) = \int_{0}^x\!\!d\xi\;k(\xi)\;,
$$
and $Y(2\pi) = 2\pi a [I_0(\epsilon) - h I_1(\epsilon)]$;  
$I_0(x), I_1(x)$ being the modified Bessel function of the First Kind for $n=0$ and $1$
respectively \cite{arfken05}.
Expression \eqref{eq:tauRA} has the advantage to make 
the $\omega$-dependence of $\tau$ explicit providing the very final expression of the 
average translocation time ready to be used in numeric computations and qualitative 
analysis.

\begin{figure}[h]
\centering
  \includegraphics[height=5cm]{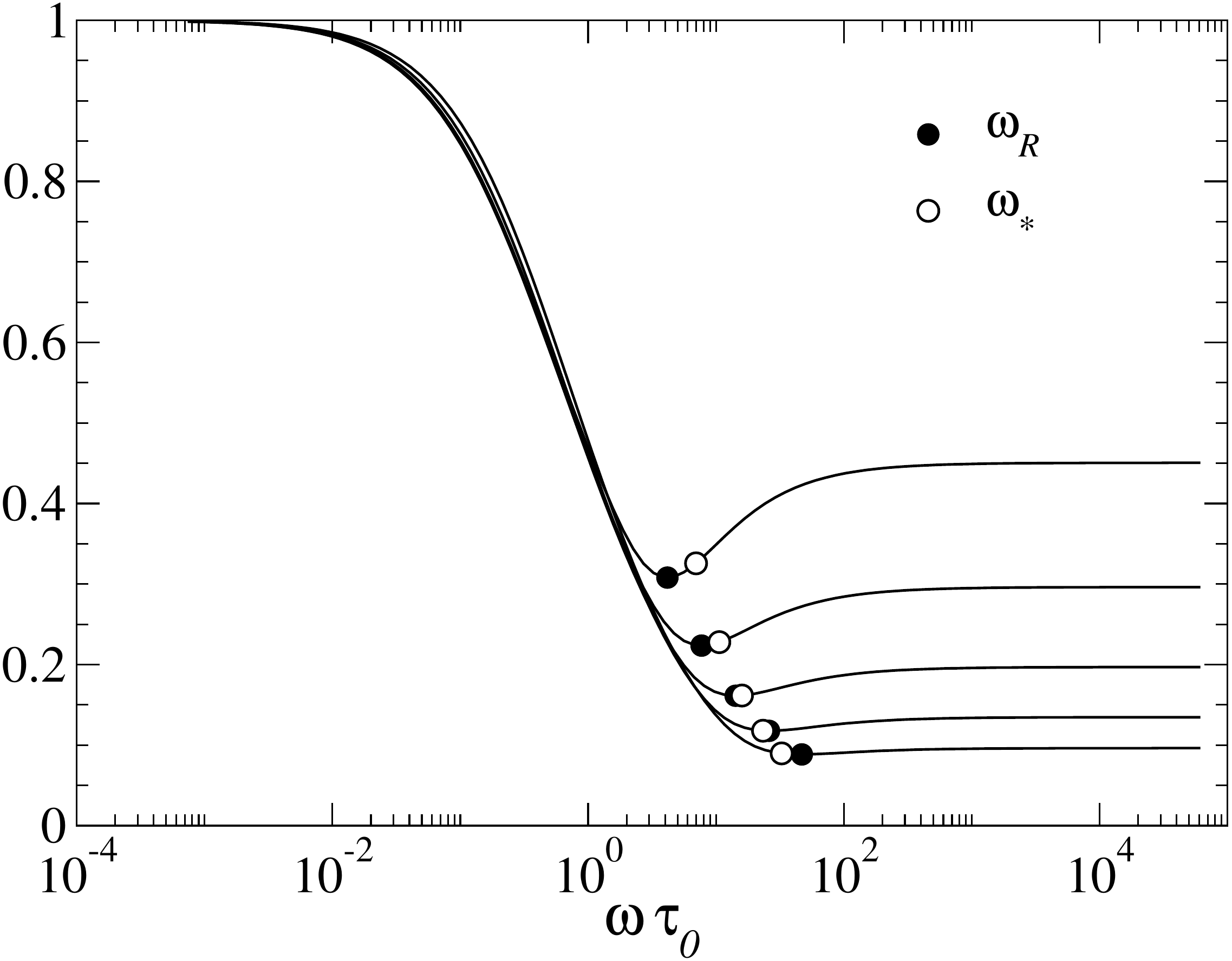}
\caption{\label{fig:demoRA} Plot of $\tau(\omega)$ from 
Eq.\eqref{eq:tau_final}, computed for the same parameters of 
Fig.\eqref{fig:two_state}, giving $\epsilon=2.4$, at  
5 values of the phase: $\phi = 2.57 + n\times 0.2$, 
$n=1,\ldots,5$. The curves clearly show the resonant minimum as 
expected from a RA process.  
Full circles mark the true minima, open circles represent the 
corresponding values of $\omega_*$, Eq.\eqref{eq:star}, 
which can be considered a rough estimate of the resonant frequency.
}  
\end{figure}

A first observation stems from the two limiting behaviours:  
$\omega\to 0$ (quasistatic barrier) yields $\tau(0) = \tau_0$, while, for 
$\omega$ large (fast oscillations), we have $\tau(\infty) = Y(2\pi) = 2\pi k_{\mathrm{eff}}^{-1}$,
where
\begin{equation}
k_{\mathrm{eff}} =  
\dfrac{e^{\epsilon\sin(\phi)}}{\tau_0\;[1+h\sin(\phi)]} 
[I_0(\epsilon) - hI_1(\epsilon)]
\label{eq:keff}
\end{equation}  
corresponds to the average of the rate \eqref{eq:ktime} over one oscillation period 
$T$, therefore we can re-write 
\begin{equation}
\tau(\omega) = \dfrac{\omega^{-1}}{1-\exp[-2\pi\;k_{\mathrm{eff}}/\omega]} 
\int_{0}^{2\pi}\!\!dx\;
\exp\bigg\{-\dfrac{Y(x)}{\omega}\bigg\}
\label{eq:tau_final}
\end{equation}
which is exactly Eq.\eqref{eq:tau_res} of Sec.\ref{sec:flupore}. 
The function $\tau(\omega)$ develops a minimum for a ``resonant'' 
frequency $\omega_R$ (see Fig.\ref{fig:demoRA}) that can be computed by 
solving numerically the equation  $\partial_{\omega}\tau=0$,   
full dots in Fig.\ref{fig:demoRA}.  

However, to arrive at an analytical expression useful for qualitative analysis,  
we can observe that $\omega_R$ is close to the crossover point $\omega_*$
determined by the condition $2\pi\;k_{\mathrm{eff}}/\omega_* \simeq 1/2$  
for the exponential at denominator of Eq.\eqref{eq:tau_final}. 
\begin{equation} 
\omega_* = \omega_0 \dfrac{e^{\epsilon\sin(\phi)}}{2[1+c\sin(\phi)]} 
[I_0(\epsilon) - hI_1(\epsilon)]\;. 
\label{eq:star} 
\end{equation} 
This value can be considered a rough yet reasonable estimate of $\omega_R$, as it 
shown in Fig.\ref{fig:demoRA}, open circles.     

\section{Toy model\label{app:Toy}}
This appendix shows the derivation of formula \eqref{eq:pdf_time} 
for the translocation time PDF.
 
The collective coordinate $Q$ is supposed to follow the driven-diffusion 
dynamics \cite{Lubensky,Berezhkovskii2003,MuthuBook,Ammenti2009}
\begin{equation}
\dot{Q} = \mu_0 F + \sqrt{2D_0}\;\eta(t)
\label{eq:LangQ}
\end{equation}
$\eta$ indicates a Gaussian noise, 
with $\langle\eta(t)\rangle =0$ and autocorrelation 
$
\langle\eta(t)\eta(s)\rangle = f(\omega t)\,\delta(t-s)
$
according to Eq.\eqref{eq:modula}, and  
$\mu_0, D_0$ are the mobility and diffusivity of $Q$ when 
$f(\omega t) = 1$, respectively. 

The solution to the stochastic differential Eq.\eqref{eq:LangQ} 
with initial condition $Q(0)=0$ is 
\begin{equation}
Q(t) = \mu_0 F t + \sqrt{2D_0}\int_{0}^t\!\! ds\;\eta(s)
\label{eq:Qsol}
\end{equation}
it defines a Gaussian process with average and spreading 
\begin{equation} 
\begin{aligned} 
& \langle Q(t) \rangle = \mu_0 F t \\
& \langle\,[Q(t) - \langle Q(t)\rangle]^2\,\rangle  = 2 D_0 \Delta(t) 
\end{aligned} 
\end{equation} 
where thanks to the $\delta$-correlation property of the noise, we have
\begin{equation}
\Delta(t) =  \int_{0}^t\!\! ds\;f(\omega s). 
\label{eq:Delta}
\end{equation}

From the solution \eqref{eq:Qsol}, it is easy to derive the coefficients 
\begin{equation}
\begin{aligned} 
& \lim_{h\to 0}\;\dfrac{\langle Q(t+h) - Q(t)\rangle}{h}  =  \mu_0 F \\
& \lim_{h\to 0}\;\dfrac{\langle\,[Q(t+h) - Q(t)]^2\,\rangle}{2h} =  D_0 f(\omega t)
\end{aligned} 
\end{equation}
of the corresponding Smoluchowski equation 
\begin{equation}
\frac{\partial P}{\partial t} = 
-\mu_0 F \frac{\partial P}{\partial Q} + 
D_0 f(\omega t)\frac{\partial^2 P}{\partial Q^2}\,,  
\label{eq:app_FPQ}
\end{equation}
which admits the fundamental solution
$$
G(Q,t) = \frac{1}{\sqrt{4\pi D_0\Delta(t)}}
\exp\bigg\{-\frac{(Q - \mu_0 F t)^2}{4 D_0\Delta(t)}\bigg\}
$$
satisfying the initial conditions $G(Q,0) = \delta(Q)$. 

$G(Q,t)$ is easily obtained via the change of variables 
$\xi = Q-\mu_0 F t$ and $\tau = \Delta(t)$ that transforms Eq.\eqref{eq:app_FPQ} 
into an equation with constant diffusivity, no drift but same initial 
condition.

As discussed in sec.\ref{sec:flupore}, the PdF of translocation times can be derived 
from the First Passage Theory (FPTh) of the process $Q(t)$ by 
assuming an initial condition $P(Q,0) = \delta(Q)$, a no-flux boundary at $Q=0$,
$J(0,t) = [\mu_0 F P - D_0 f(\omega t) \partial_Q P]_0 = 0$ and an absorbing 
boundary at $Q=1$, $P(1,t) = 0$.    

The exact solution to this problem is not available as it is not separable,
in addition, the boundaries introduce further complications.
However, a meaningful approximation can still be obtained by 
assuming that the drift is strong enough to induce a swift displacement of the process 
from the $0$-boundary, hence, the effect of the impenetrable barrier at $Q=0$ 
becomes negligible and, in a first approximation, it can be shifted to $Q = -\infty$. 
 
The other boundary instead is fulfilled trying to ``extend'' the method of images 
to the case of time-dependent diffusion coefficients as shown in Ref.\cite{molini2011}.
Accordingly, we attempt the solution 
\begin{equation}
P(Q,t) \simeq  G(Q,t) - w G(Q-2,t)
\label{eq:Image}
\end{equation}
originated from the new initial condition $P(Q,0) = \delta(Q) + w \delta(Q-2)$,  
where the extra term represents an auxiliary symmetric source (the image) with 
respect the boundary $Q=1$. 

The coefficient $w$ is to be determined by imposing the boundary 
condition $P(1,t) = 0$ and leading to 
$w = G(1,t)/G(-1,t) = \exp\{\mu_0 F t/[D_0\Delta(t)]\}$. 

In Eq.\eqref{eq:Image} we did not use the ``='' to stress that it is  
only a ``pseudo''-solution, this can be verified immediately 
by plugging it back to Eq.\eqref{eq:app_FPQ}. 

When the approximate solution $P(Q,t)$ is plugged into Eq.\eqref{eq:psi} and 
in the following ones we obtain the final analytical form \eqref{eq:pdf_time} 
for the translocation time PdF.   

\begin{figure}
 \centering
\includegraphics[height=5cm]{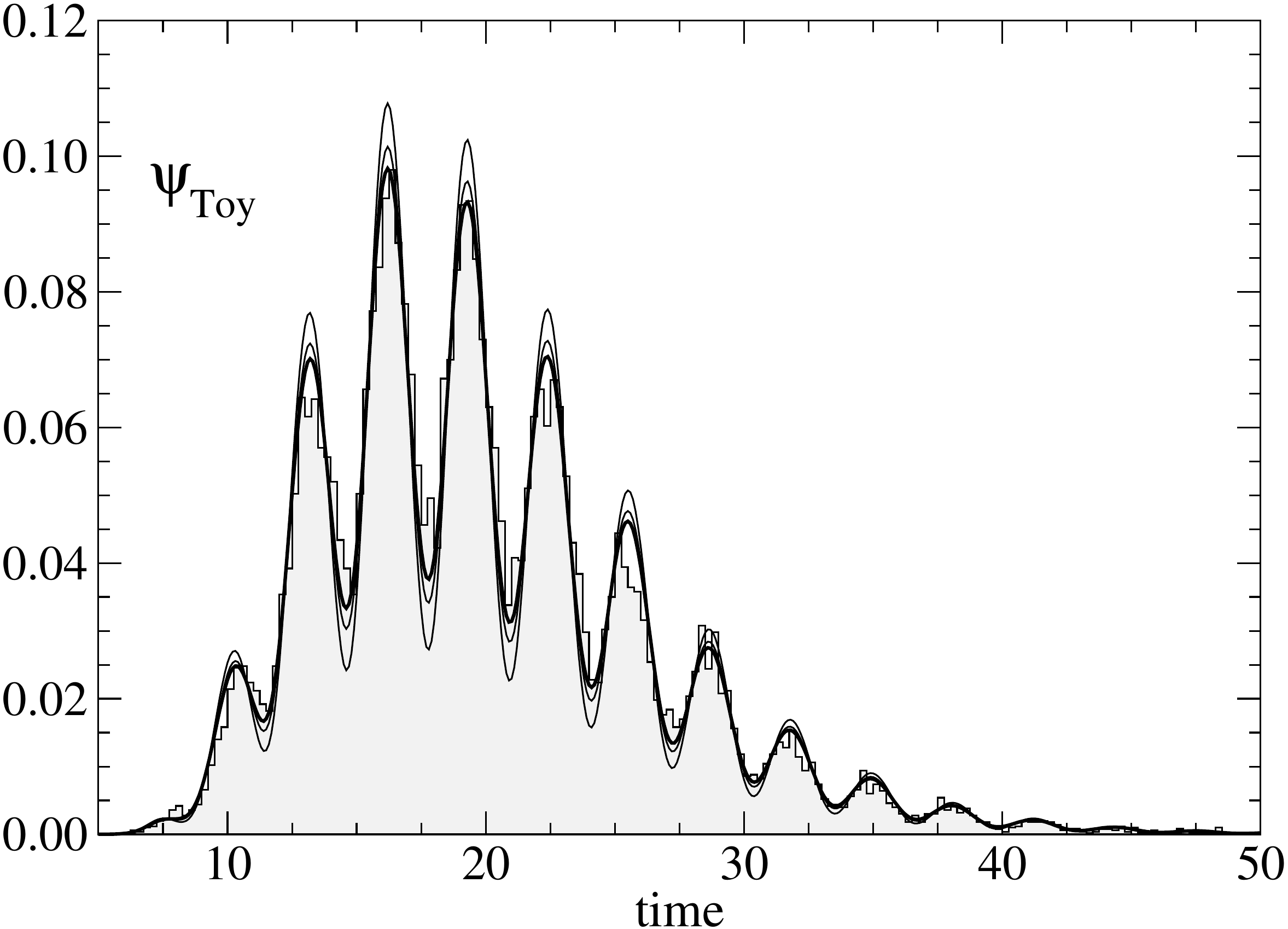}
\caption{\label{fig:test} Test of toy model: comparison of the translocation 
time PdF Shaded region is the histogram of the First Exit Time form the boundary 
$Q = 1$ with the corresponding theoretical PdF \eqref{eq:pdf_time} from the simple 
driven-diffusion model of parameters:
$\mu_0=1, D_0 = 1, F=0.05$, $h = 0.60$ and $\phi = 0$.
The different lines represent plots of \eqref{eq:pdf_time} at different 
values of $h$ showing that this parameter can be adjusted to obtain an optimal
match of the approximated theory and simulations data.}
\end{figure}

Since the theoretical distribution \eqref{eq:pdf_time} is not exact 
for the model \eqref{eq:LangQ}, it is important to test its accuracy against the 
true PdF of exit times that can be computed by a direct numerical integration 
of Eq.\eqref{eq:LangQ}. 
Fig.\ref{fig:test} shows the reasonable match of $\psi(t)$
with the histogram of first exit times from the boundary $Q = 1$, obtained by a 
numerical integration of Eq.\eqref{eq:LangQ}. 
The different curves show that the accuracy is improved by choosing an optimal value 
of $h$.

\bibliographystyle{rsc}
\bibliography{biblio}

\end{document}